\begin{filecontents}{pictexa.tex}
\documentclass{article}
\usepackage{epic,eepic,pspicture,mathptmx,times,graphpap}
\documentclass{cernrep}
\def\ttbar{$t\overline{t}~$}  
\def\bbbar{$b\overline{b}~$} 
\def\pbarp{$p\overline{p}~$}             
\def\qqbar{$q\overline{q}~$} 
\begin{document}
\title{LHC Highlights and Prospects}
\author{Cecilia E.\ Gerber}
\institute{University of Illinois at Chicago, Chicago, IL, USA}

\begin{abstract}
These lectures were presented at the 2019 CERN Latin-American School of High Energy Physics. They were centered on the experimental methods used in hadron colliders to advance our understanding in the field of high energy particle physics. 
From accelerators, to particle detector technologies, object identification and data analyses techniques, the lectures did not attempt to provide a comprehensive, in-depth technical background, but rather focused on an overview of experimental techniques that enabled our advances in supporting and challenging the predictions of the standard model. This document includes a selection of the material presented in the lectures, focusing on how advances in detector technologies and object identification enabled the development of increasingly sophisticated data analysis techniques. This writeup also includes an outlook to the future LHC program and beyond. 
\end{abstract}

\keywords{CERN report; LHC, Hadron Colliders}

\maketitle

\section{Introduction}
The Large Hadron Collider~\cite{lhc}, a $26.7\;\rm km$ diameter superconducting proton-proton collider, is the largest particle accelerator ever built. It is operated by the European Organization for Nuclear Research (CERN) in Geneva Switzerland, and is the last stage of a multi-accelerator complex that results in proton-proton collisions at a center of mass energy in the $7 - 14\;\rm TeV$ range. A schematic view of the CERN accelerator complex can be seen in figure~\ref{figlhc}. The first stage is the LINAC~2 linear accelerator, where a proton source extracts $90\;\rm keV$ protons from a Hydrogen bottle and accelerates the beam to $50\;\rm MeV$ over a distance of $33\;\rm m$, providing a pulse every $1.2\;\rm s$. Next, the PS Booster, which is the first synchrotron in the accelerator chain, and has $157\;\rm m$ in circumference, increases the proton energy to $1.4\;\rm GeV$ in $1.2\;\rm s$. Protons are then injected into the PS, the oldest operating synchrotron at CERN with a circumference of $628\;\rm m$ (4 times the size of the PS Booster), and their energy is increased to $26\;\rm GeV$. Next, protons are injected into the SPS, the first superconducting synchrotron in the chain, with $6.9\;\rm km$ circumference and $30\;\rm m$ underground, which was originally a proton-antiproton collider that lead to the discovery of the $W$ and $Z$ bosons. The SPS increases the proton energy to 450~GeV and provides beam to the LHC and to fixed target areas. The LHC consists of 1232 main dipoles of 15~m each that deviate the beams around the 27~km circumference, 858 main quadrupoles that keep the beam focused and 6000 corrector magnets to preserve the beam quality. The main magnets use superconducting cables (Cu-cladded Nb-Ti), with 12,000~A providing a nominal field of 8.33~Tesla. The LHC started operations in 2010, delivering $36\;\rm pb^{-1}$ of data at $\sqrt{s}=7\;\rm TeV$, followed by $5\;\rm fb^{-1}$ of data in 2011. The center of mass energy was increased to $\sqrt{s}=8\;\rm TeV$ in 2012, when $20\;\rm pb^{-1}$ of data were delivered. Finally, in what is known as Run 2, $150\;\rm pb^{-1}$ of data at $\sqrt{s}=13\;\rm TeV$ were delivered between 2015 and 2018.

\begin{figure}[ht]
\begin{center}
\includegraphics[width=15cm]{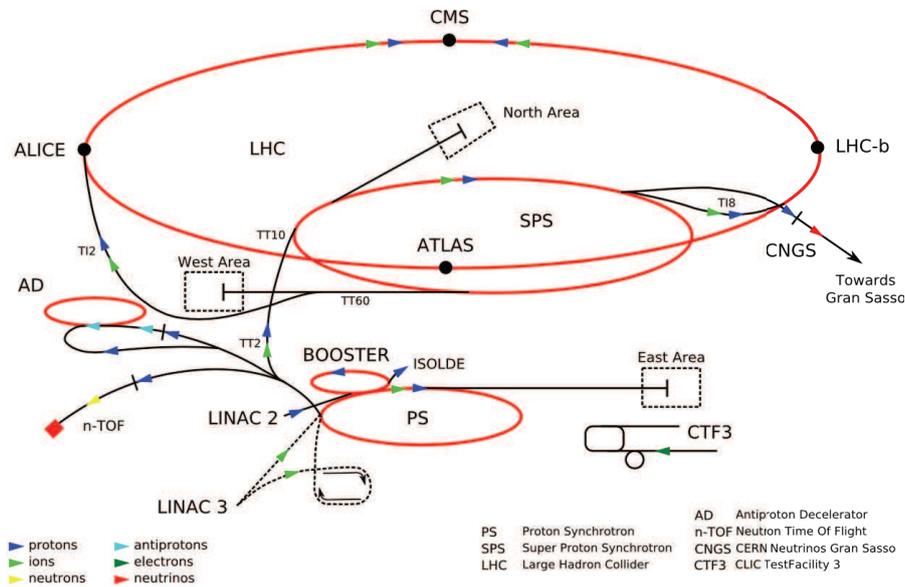}
\caption{Schematic view of the CERN accelerator complex, showing the different accelerator stages that result in proton-proton collisions at the center of the four detector areas.}
\label{figlhc}
\end{center}
\end{figure}

These lectures will focus only on the two multi-purpose experiments, ATLAS~\cite{ATLAS} and CMS~\cite{CMS}, shown schematically in figures~\ref{atlas} and~\ref{cms}. Both detectors rely on layered instrumentation technologies to detect the passage of particles and measure their position and their energies with the best possible precision. The innermost layer is typically a silicon-based semiconductor tracker, that is both radiation hard and has low mass, to minimize the multiple scattering from detected tracks. Electrically charged particles leave hits in the tracker layers, which allows trajectories to be reconstructed from consecutive measurements as particles traverse the detector volume.  CMS is the first hadron collider experiment to use an all-silicon tracker. ATLAS innermost tracker is silicon-based, but the outer part uses a gas and wire transition radiation tracker. Outside of the trackers are the calorimeters, that destructively measure the energy of charged and neutral particles. Calorimeters complement the information obtained from the magnetic spectrometers, as they are able to measure the energy of neutral particles and have an energy resolution which improves with the particle energy, while the spectrometer resolution degrades with particle energy. Finally, the outermost part of the detectors are the muon spectrometers, that are typically gas-based detectors that cover huge volumes and identify the only particles that escape the calorimeters, which are the muons. Both collaborations have superconducting magnets. CMS has a single solenoidal 4~Tesla magnet, cooled at -270~C, outside the combined tracker and calorimeter volume. The muon chambers are interleaved with a 12-sided, 3-layers iron structure that surrounds the magnet coils and contains the B field. ATLAS has a 2~Tesla central solenoid between the tracker and the calorimeters and two toroids (a central and a forward toroid) as part of the ATLAS muon system. Finally, the trigger systems decide, in real-time, which subset of data is to be readout by the detector and stored for offline analysis, and the Data AcQuisition (DAQ) system collects the data from the different parts of the detector, converts it to a suitable format, and saves it to permanent storage. Both detectors use a multi-level trigger system, with a first trigger level with very short latency, high signal efficiency but modest background rejection, which is typically  firmware-based, single-detector or limited combination. This is followed by subsequent trigger level(s) that achieve high background rejection, with typically larger latency, and which are computer-based and use the information from multiple detectors, with the highest level being a speed-up version of the offline reconstruction. 

The latest development in offline event reconstruction uses the particle flow algorithm~\cite{pf} to create a comprehensive list of all final-state particles in the collision. This approach makes use of the best combination of all subdetectors to reconstruct and identify all particles, and also provides robust handles against energy deposits originating from the underlying event and from multiple proton-proton interactions in the same bunch crossing. It also opens the door to a new field of studies that uses jet substructure techniques to differentiate between quark and gluon jets based on the differences in their radiation patterns and lifetimes~\cite{tag}. The technique can also be extended to distinguish jets from hadronic decays of high transverse momentum heavy particles, in particular $W$ and $Z$ bosons, top quarks and Higgs bosons. As the largest branching ratios for these particles is into their hadronic decays, these tools open up a large amount of acceptance that was previously unaccessible. Furthermore, the ability to reconstruct subjets from merged decays can be used to separate these heavy objects from the much more copious quantum chromodynamics (QCD) multijet production. This is particularly important when studying hadron collisions that give raise to a wide variety of processes with production cross sections that span 12-13 orders of magnitude. While they enable a rich physics program, the interesting processes are overwhelmed by mundane processes that occur at much higher rates. Background discrimination and residual background modeling are crucial ingredients in any physics analysis using hadron collider data. Figure~\ref{xsec} shows the cross sections and events per second produced at hadron colliders as a function of the center of mass energy.

\begin{figure}[ht]
\begin{center}
\includegraphics[width=15cm]{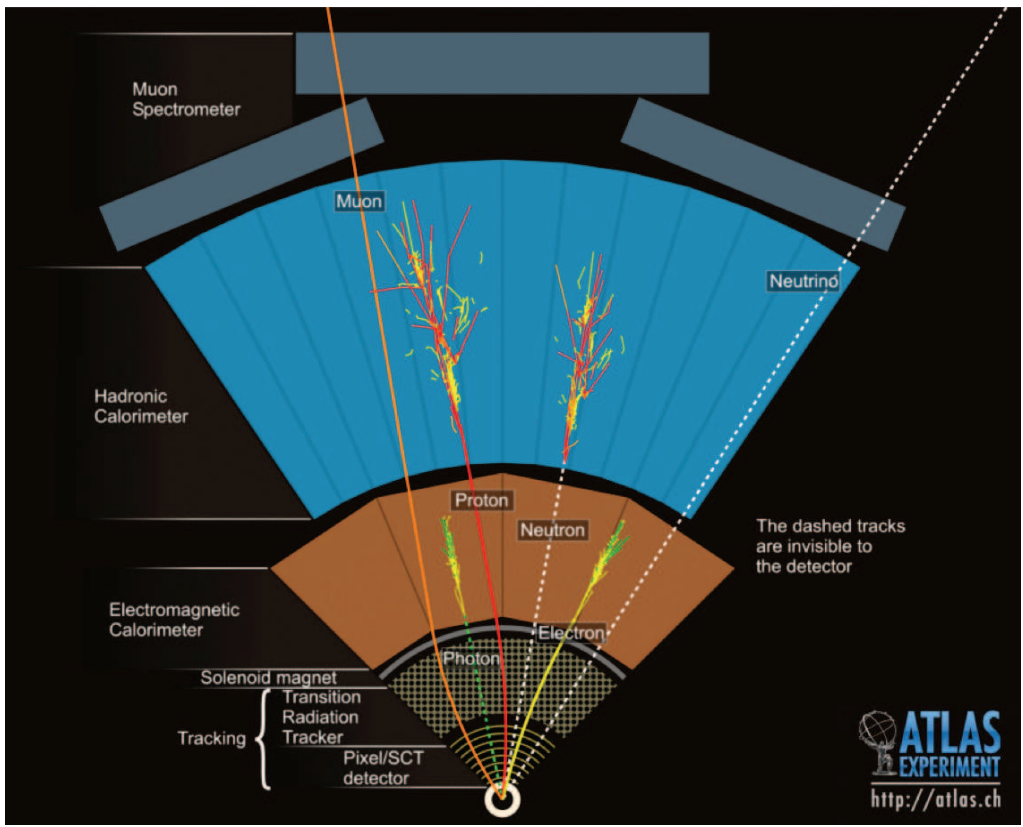}
\caption{Cross-sectional view of the ATLAS detector.}
\label{atlas}
\end{center}
\end{figure}

\begin{figure}[ht]
\begin{center}
\includegraphics[width=15cm]{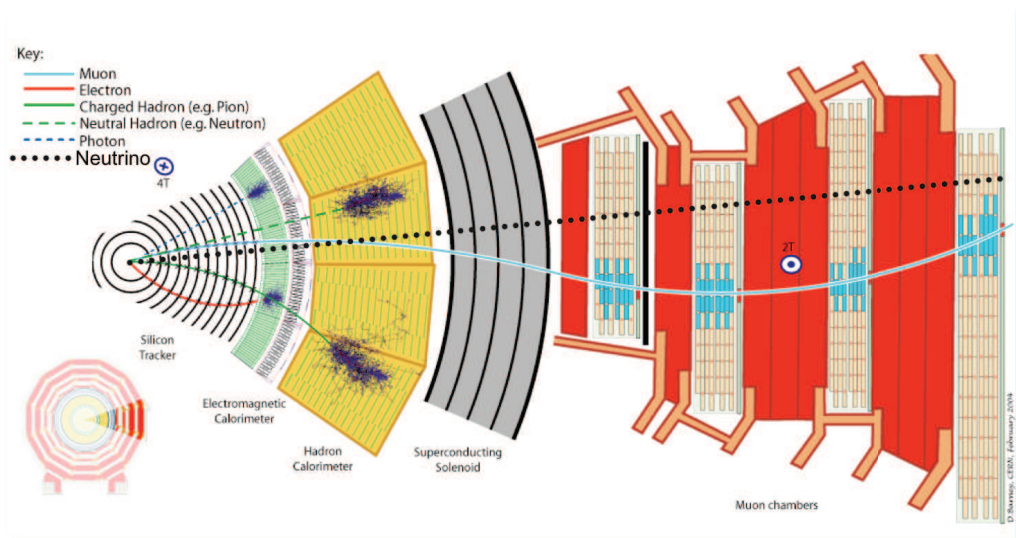}
\caption{Cross-sectional view of the CMS detector.}
\label{cms}
\end{center}
\end{figure}

\begin{figure}[ht]
\begin{center}
\includegraphics[width=15cm]{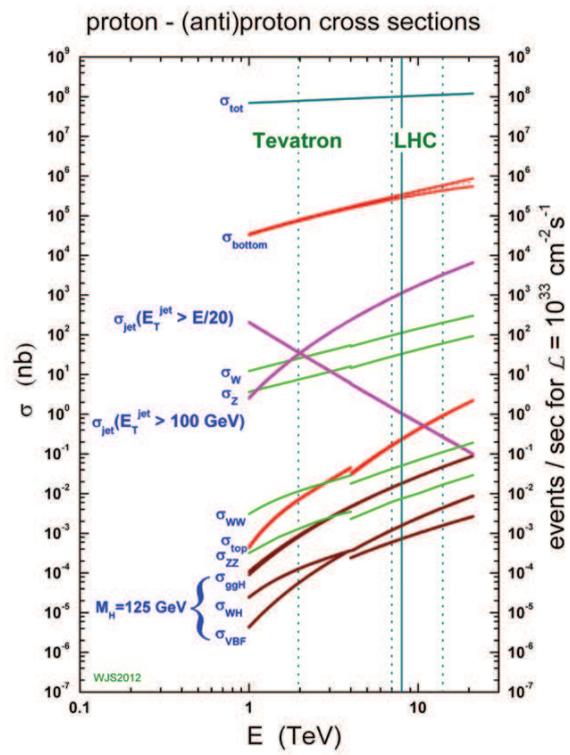}
\caption{Production cross sections as a function of center of mass energy in hadron colliders.}
\label{xsec}
\end{center}
\end{figure}

\clearpage
\section{Jet Production}
At LHC energies, the dominant process in pp collisions is jet 
production. Within the framework of QCD, inelastic 
scattering between two protons can be described as an elastic 
collision between a single constituent of each proton. These constituents are called partons. After the 
collision, the outgoing partons manifest themselves as localized streams of 
particles referred to as ``jets''. Theoretical predictions for jet production 
are given by the folding of the parton scattering cross sections with 
experimentally determined parton density functions (pdf's). These 
predictions are known at next--to--leading order (NLO) in perturbative QCD 
scattering calculations~\cite{jet1, jet2} and accurately measured 
pdf's~\cite{jet3}. Some of the questions that can be addressed with 
studies of jet production are testing of NLO QCD, extraction of pdf's,
measuring the value of the strong coupling constant $\alpha_s$, and testing
of beyond the standard model (BSM) theories.

The simplest test that can be performed is the measurement of the production rate of jets as a function of the jet transverse momentum ($p_T$) in different rapidity bins $y$, a study in which each jet in the event corresponds to an entry in the histogram. 
The ATLAS and CMS 
collaborations measure the double differential inclusive jet 
cross section in pp collisions at $\sqrt{s}=8\;\rm TeV$, which can be expressed as:
\[ 
d^2\sigma/(dp_T dy)=(N_{Jet})/(\varepsilon\Delta p_T\Delta |y| \int L dt)\]
where $N_{Jet}$ is the total number of jets observed in a certain
 jet transverse energy $E_T$ bin, $\varepsilon$ is the selection efficiency, 
$\Delta p_T$ is the bin width, $\Delta y$ is the
rapidity range considered, and $\int L dt$ is the integrated luminosity
associated with the data set. The results are shown in figure~\ref{fig1} for ATLAS~\cite{ATLAS-jet1} and figure~\ref{fig1-2} for CMS~\cite{CMS-jet1}. The cross sections for different rapidity bins are multiplied by factors indicated in the legends for better visibility. The dominant systematic uncertainties are the jet energy scale and resolution, and range from $2-30\%$, being largest for low $p_T$ jets and jets in high rapidity regions. Overall, reasonable good agreement is observed between data and the NLO predictions. The data should provide improved constraints on parton distribution functions. 

\begin{figure}[ht]
\begin{center}
\includegraphics[width=9cm]{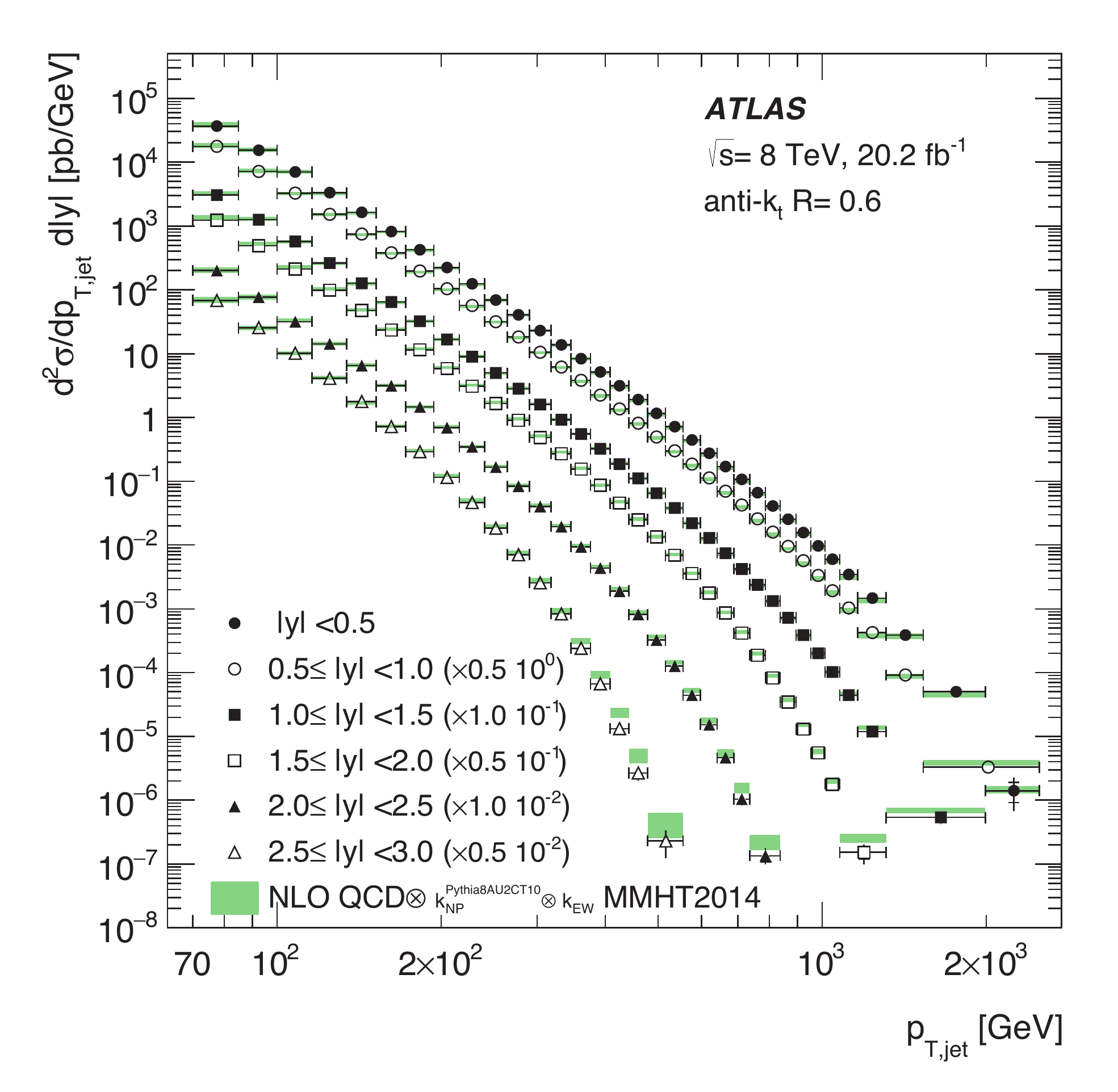}
\caption{Double-differential inclusive jet cross sections as function of jet $p_T$ from ATLAS. The cross sections for different rapidity bins are multiplied by factors indicated in the legends for better visibility.}
\label{fig1}
\end{center}
\end{figure}

\begin{figure}[ht]
\begin{center}
\includegraphics[width=9cm]{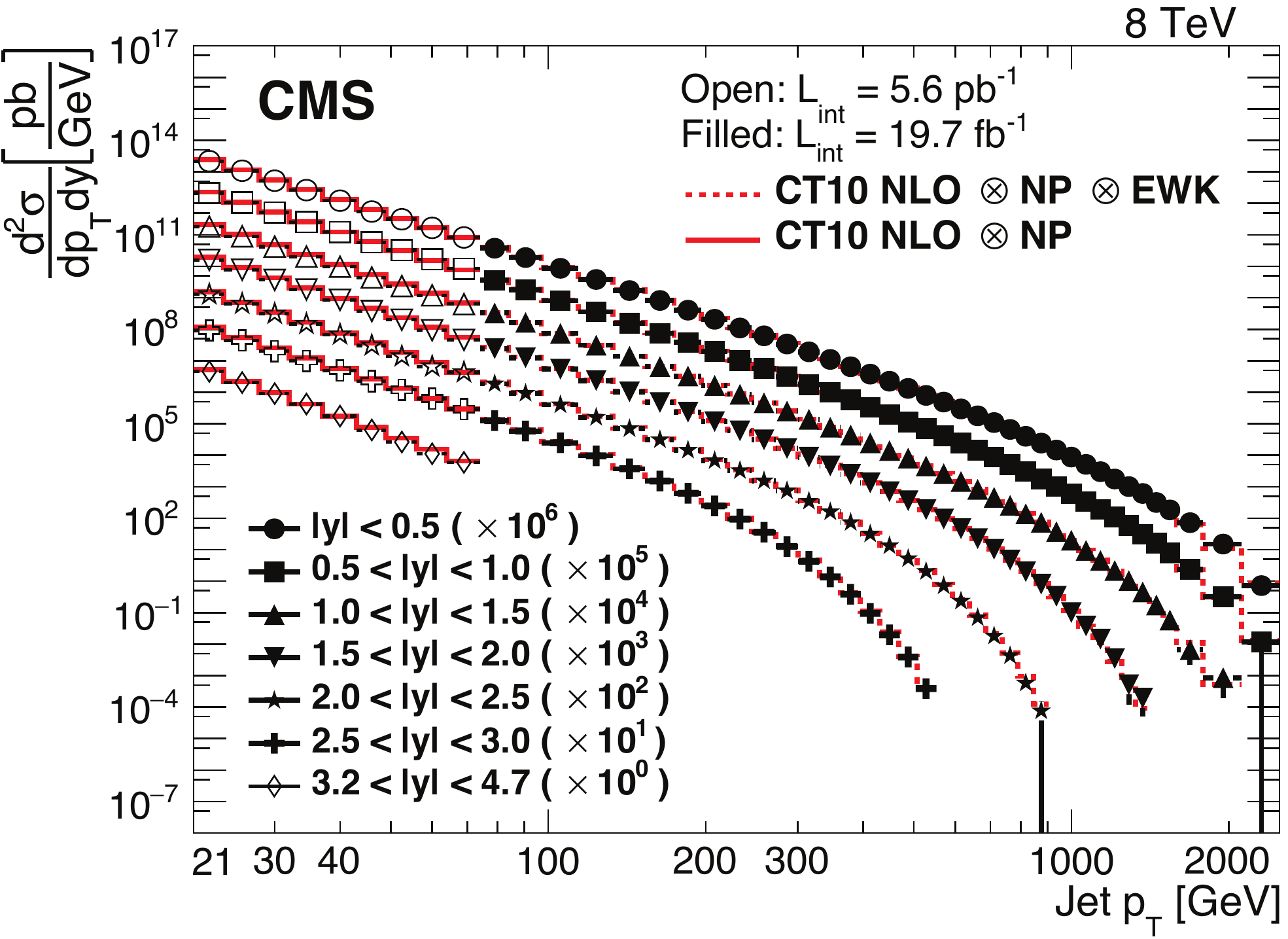}
\caption{Double-differential inclusive jet cross sections as function of jet $p_T$ from CMS. The cross sections for different rapidity bins are multiplied by factors indicated in the legends for better visibility.}
\label{fig1-2}
\end{center}
\end{figure}

Both collaboration also study the characteristics of the system given by the two leading jets in an event. Discrepancies with QCD predictions could indicate beyond the standard model physics like quark compositeness, excited quarks, quark contact interactions, extra spatial dimensions, quantum black holes, or dark matter. The results are shown in figure~\ref{fig2}. ATLAS presents its result as the dijet cross section as a function of the dijet invariant mass $m_{jj}$. CMS chooses to 
concentrate on the angular distribution of dijets relative to the beam direction by studying $\chi_{\rm dijet} = e^{|y_1 - y_2|}$, where $y_1$ and $y_2$ are the rapidities of the two leading jets. The choice of $\chi_{\rm dijet}$ as the sensitive variable is motivated by the fact that BSM processes, that are expected to have isotropic angular distributions, 
would result in an excess of events over QCD predictions at low values of $\chi_{\rm dijet}$.

\begin{figure}[ht]
\begin{center}
\includegraphics[width=9cm]{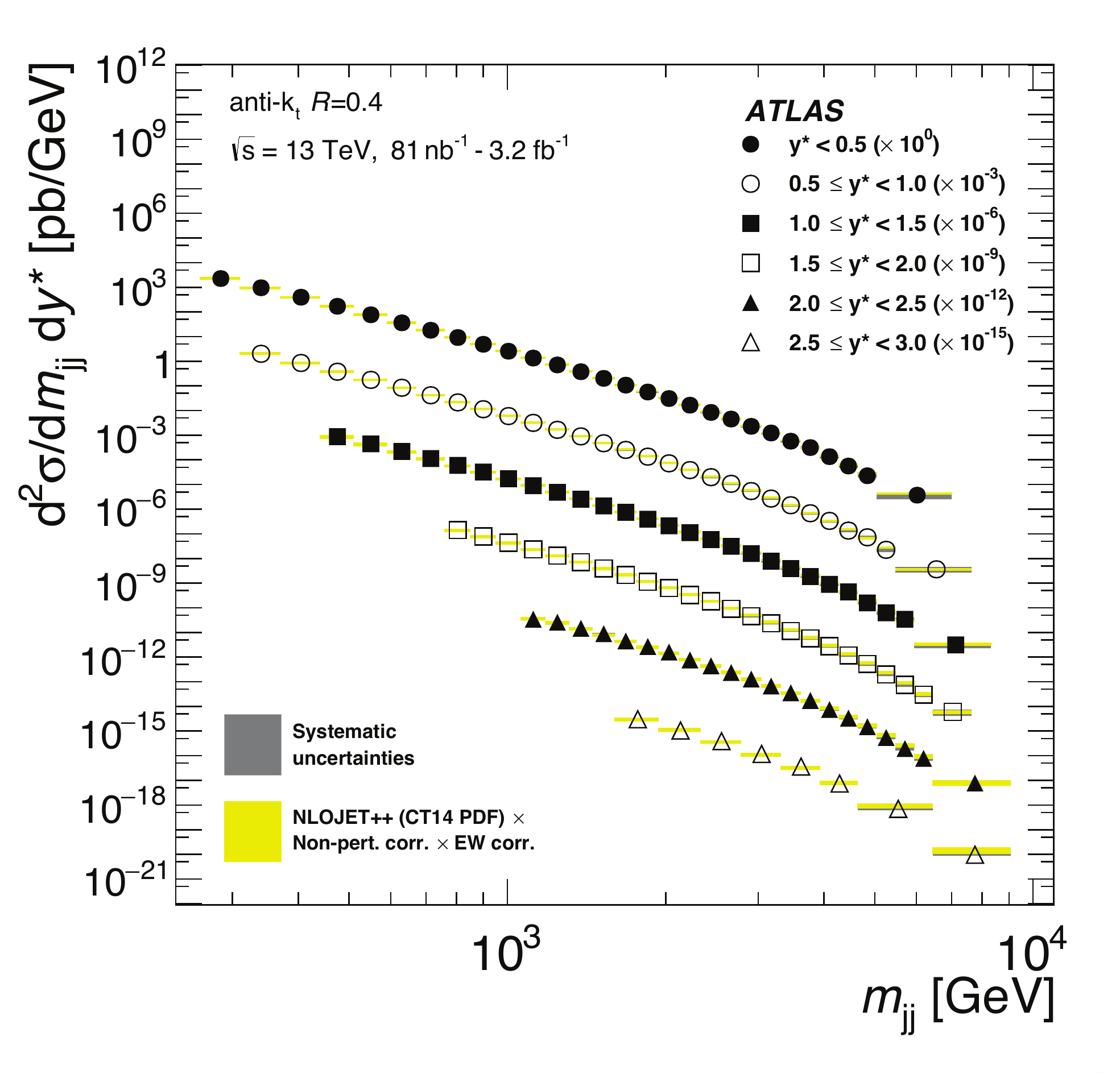}
\includegraphics[width=9cm]{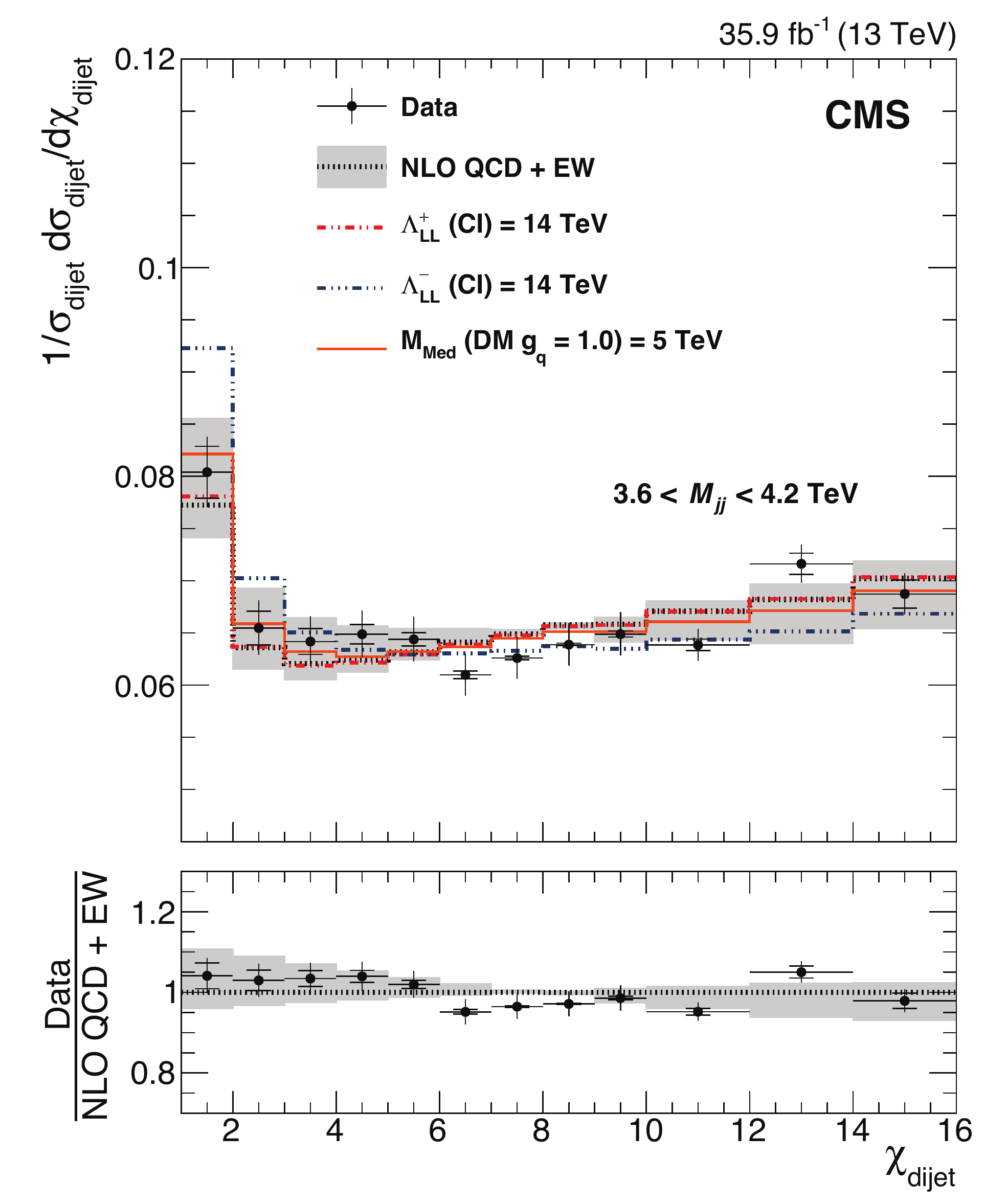}
\caption{Dijet cross-sections as a function of the dijet invariant mass $m_{jj}$. The ATLAS data are compared to NLO pQCD predictions (top). Example of a normalized 
$\chi_{\rm dijet}$ distribution where unfolded data from CMS is compared to NLO predictions and various BSM scenarios (bottom).} 
\label{fig2}
\end{center}
\end{figure}

\clearpage
\section{Vector Boson Production}
$W$ and $Z$ bosons, the carriers of the weak force, are directly produced in
high energy pp  collisions at the CERN LHC. 
In addition to probing electroweak physics, the study of the
production of $W$ and $Z$ bosons
provides an avenue to explore QCD. $W$ and $Z$ bosons, when produced in association with jets, and in particular with $b$-jets, constitute an irreducible background to many processes that decay to $W$ or $Z$, like top, Higgs and BSM production. The leptonic $Z$ decays provide clean samples with adequate statistics for detector performance measurements and the extraction of data-to-MC correction factors for trigger and lepton identification.

The most precise measurement of the leptonic $W$ production cross section at $\sqrt{s}=8\;\rm TeV$ was performed by CMS using special low pileup data collected in 2012~\cite{CMS-VB}. The data sample corresponds to an integrated luminosity of $18.2\pm0.5\;\rm pb^{-1}$, and has an average of 4 interactions per bunch crossing, to be compared with an average of 21 interactions for the regular beam conditions during 2012. Events are selected in their decays to high $p_T$, isolated electrons or muons. The leading systematic uncertainty arises from the measurement of the lepton reconstruction and identification. The uncertainty on the integrated luminosity cancels when calculating the ratio of cross sections, which are measured with a precision of $2\%$. Figure~\ref{fig3} shows the measured and predicted total production cross sections times branching ratio to leptons for $W$ vs. $Z$ (left) and $W^+$ vs $W^-$ (right). The measurements in the electron and muon channel are in agreement with NNLO theoretical predictions and among the channels, in accordance with the expectation from lepton universality.

\begin{figure}[ht]
\begin{center}
\includegraphics[width=7.5cm]{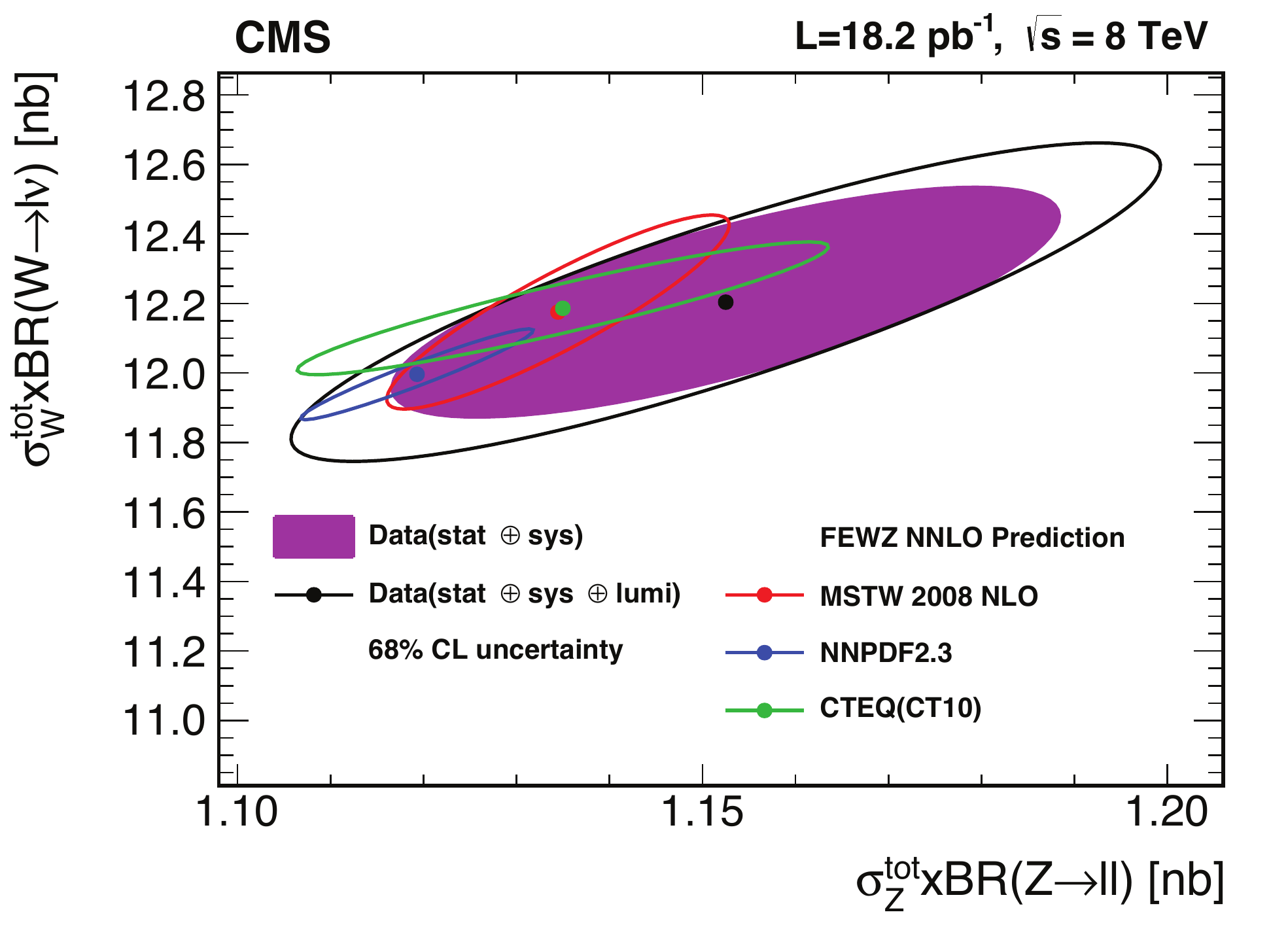}
\includegraphics[width=7.5cm]{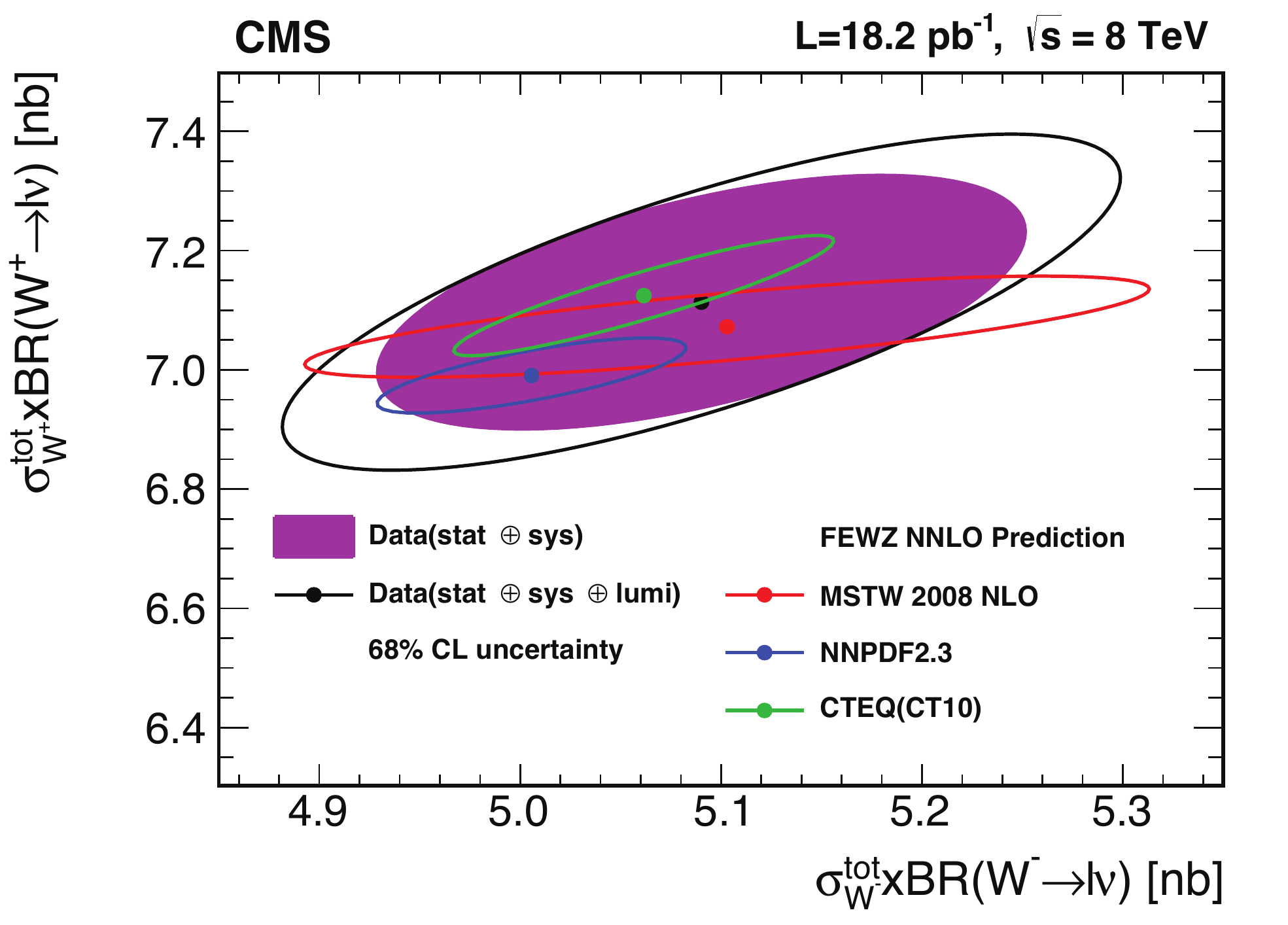}
\caption{Total production cross sections times branching ratio to leptons for $W$ vs. $Z$ (left) and $W^+$ vs $W^-$ (right). The ellipses illustrate the $68\%$ CL coverage for total uncertainties (open) and excluding the luminosity uncertainty (filled). The uncertainties in the theoretical predictions correspond to the PDF uncertainty components only and are evaluated for different PDF sets, as indicated in the figure.} 
\label{fig3}
\end{center}
\end{figure}

\clearpage
\section{Measurement of the $W$ mass}
In the standard model of the electroweak interactions, the mass
of the $W$ boson can be expressed as 
\begin{equation}
m^{2}_{W}   \bigg(1 - \frac{m^{2}_{W}}{m^{2}_{Z}}\bigg) = \frac{\pi\alpha(m_Z^2)}
 {\sqrt{2}G_F} (1+\Delta r_W)\; .
\label{eq:dr}
\end{equation}
A measurement of $m_W$,
together with $m_Z$, the Fermi constant ($G_F$), and the electromagnetic
coupling constant ($\alpha$),
determines the electroweak radiative corrections $\Delta r_W$ experimentally.
The dominant contributions to $\Delta r_W$
arise from loop diagrams that involve the top quark and the Higgs boson.
The correction from the $t\overline b$ loop is substantial
because of the large mass difference between the two quarks. It is proportional
to $m_t^2$ for large values of the top quark mass $m_t$. If additional particles which couple to the $W$ boson exist, they would give
rise to additional contributions to $\Delta r_W$. Therefore, a measurement of
$m_W$ is one of the most stringent experimental tests of SM predictions.
Deviations from the predictions may indicate the existence of new physics.
Within the SM, measurements of $m_W$ and the mass of the top quark and the Higgs boson are a crucial test of the overall consistency of the SM, and discrepancies could indicate new physics. The experimental challenge is thus to measure the $W$
boson mass to sufficient precision, about $0.1\%$, to be sensitive to BSM effects. 

The ATLAS collaboration has used data collected in 2011 in pp collisions at $\sqrt{s}=7\;\rm TeV$ to measure the $W$ boson mass from template fits to the reconstructed distributions for the charged lepton $p_T$ and the $W$ boson transverse mass~\cite{ATLAS-WMass}. The analysis uses $4.6\;\rm pb^{-1}$ of data, which results in $7.8\times 10^6$ $W\to\mu\nu$ and $5.9\times 10^6$ $W\to e \nu$ candidates. In addition, $1.23\times 10^6$ $Z\to \mu \mu$ and $0.58\times 10^6$ $Z\to ee$ candidates are used to pin down the lepton energy and $W$ recoil calibration. The measured mass of 
$m_W = 80370\pm7{\rm(stat.)}\pm11{\rm(exp. syst.)}\pm14{\rm(mod. syst.)\; MeV} = 80370 \pm 19\;\rm {MeV}, $
is consistent with previous results and SM expectations. The measurement is limited by the understanding of the detector modeling, the PDFs and theoretical uncertainties. 
Figure~\ref{fig4} shows the measured values for the $W$ boson mass and results from global electroweak fits. 

\begin{figure}[ht]
\begin{center}
\includegraphics[width=7.5cm]{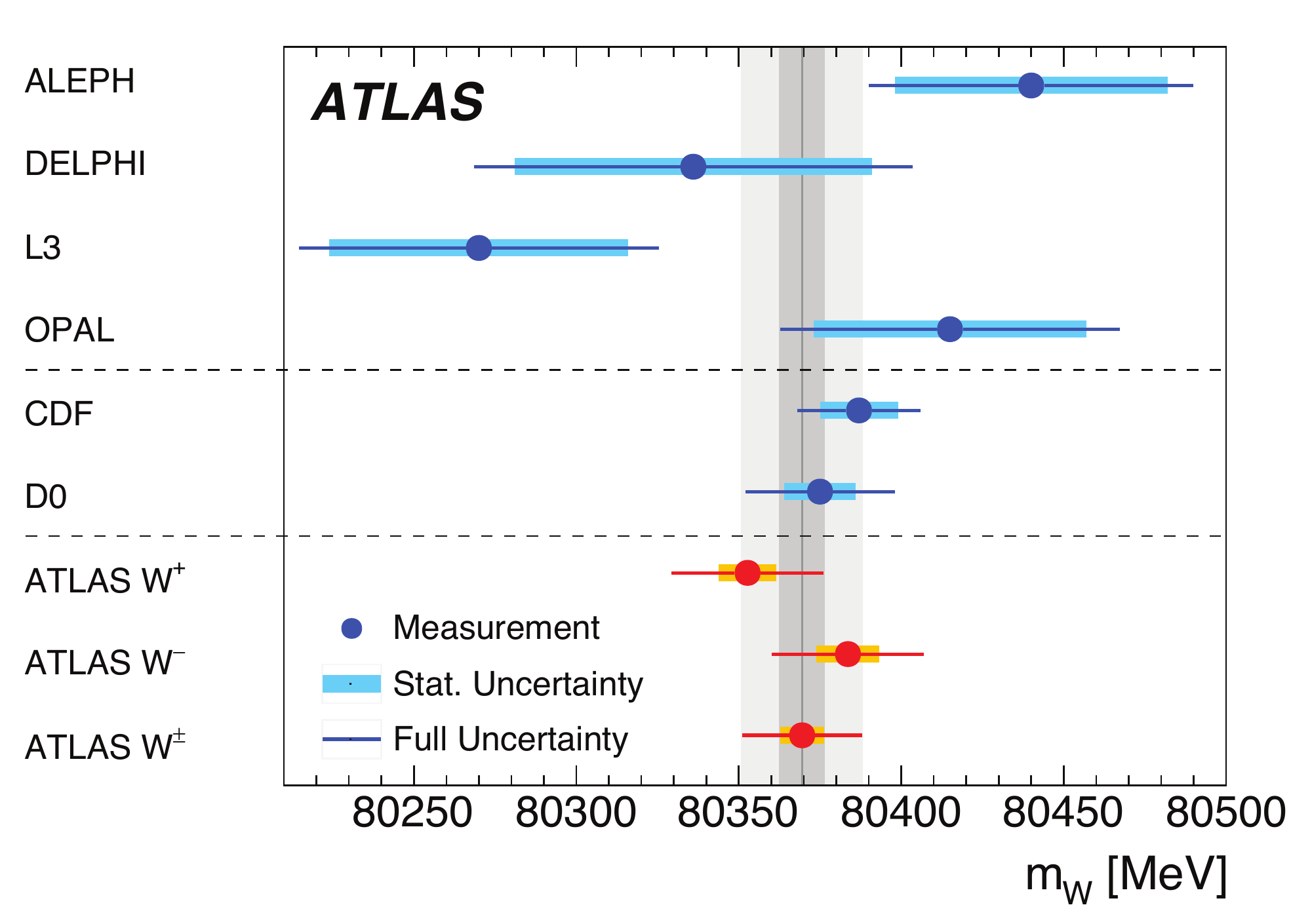}
\includegraphics[width=7.5cm]{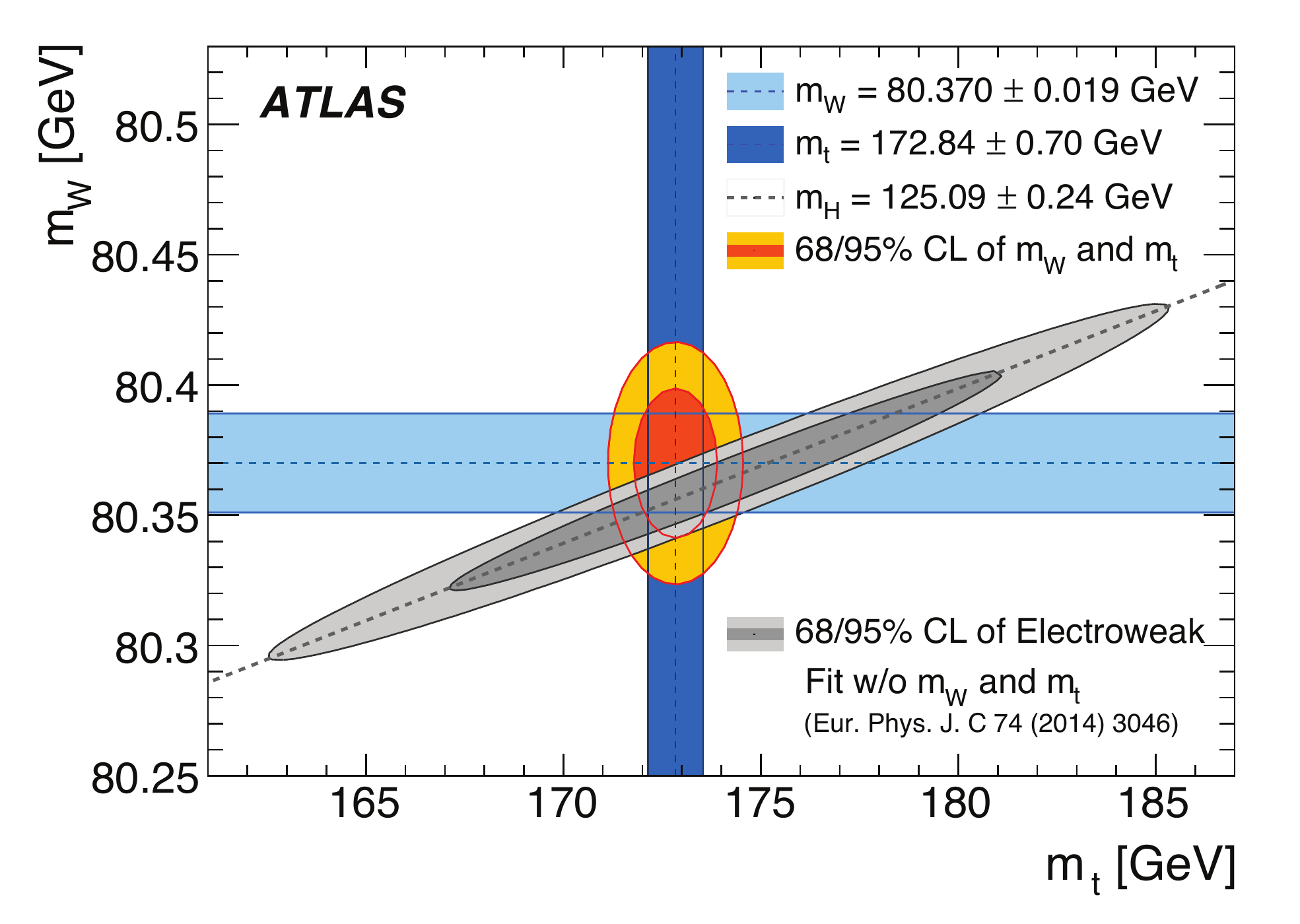}
\caption{Comparison of the measured values for the $W$ boson mass by various collaborations (left). ATLAS measured values for the mass of the $W$ boson and the top quark are compared to the results from global electroweak fits based on the measured LHC Higgs Boson mass of $125.09\pm0.24\;\rm{GeV}$ (right).}  
\label{fig4}
\end{center}
\end{figure}

\clearpage
\section{Top Quark Production and Decay}

The top quark is the heaviest known elementary particle and 
completes the quark sector of the three-generation structure of the
standard model (SM). It differs from the
other quarks not only by its much larger mass, but also by its
lifetime which is too short to build hadronic bound states.

The SM predicts that top quarks are created via two 
independent production mechanisms at hadron colliders. 
The primary mode, in which a \ttbar\ pair is produced from a $gtt$ 
vertex via the strong interaction, was used by the D0 and CDF 
collaborations to establish the existence of the top quark in 
1995~\cite{top-obs-1995-cdf,top-obs-1995-d0}. 
The second production mode of top quarks at hadron colliders is the 
electroweak production of a single top quark from a $Wtb$ vertex. 
At Tevatron energies, the predicted cross section for single top quark production is about half 
that of \ttbar\ pairs but the signal-to-background ratio is much worse; 
observation of single top quark production 
has therefore been impeded by its low rate and difficult 
background environment compared to the top pair 
production, and only detected by the same collaborations in 2009, fourteen years after the strong production~\cite{d0-obs,cdf-obs}. 

Within the SM, the top quark decays almost exclusively
into a $W$ boson and a $b$ quark, resulting in two $W$ bosons and two 
$b$ jets in each \ttbar pair event. 
The $W$ boson itself decays into one lepton and its 
associated neutrino, or hadronically. 
The \ttbar pair decay channels have been classified as follows:
the dilepton channels where both $W$ bosons decay leptonically 
into an electron or a muon, 
the lepton+jets channels where one of the $W$
bosons decays leptonically and the other hadronically, 
and the all-jets channel where both $W$ bosons decay
hadronically. \ttbar\ pair production cross sections have been measured in \pbarp collisions at the Tevatron and in pp collisions at the LHC in all
decay channels except in the dilepton channel where both Tau leptons decay hadronically. Figure~\ref{fig5} summarizes the measurements of the \ttbar production cross-section as a function of the center of mass energy compared to theoretical predictions at NNLO QCD with NNLL resummation~\cite{TOPWG}. 
Excellent agreement between measurements and theoretical predictions is observed in all channels. Precise measurements of the \ttbar production cross section represent a test of pQCD at high $Q^2$, can be used to constrain PDFs, to determine the top quark mass $m_t$ , and measure  the strong coupling constant $\alpha_s$. Precise comparisons of the measured cross sections in different channels and different methods with theoretical expectations are sensitive to new physics. Good understanding of the composition of the samples is crucial to enable the measurement of top quark properties and as input to searches for which the top quarks are the dominant backgrounds. Well understood samples of top quarks can also be used to constrain the energy scale of jets and to measure efficiencies to tag b and top jets.

\begin{figure}[ht]
\begin{center}
\includegraphics[width=15cm]{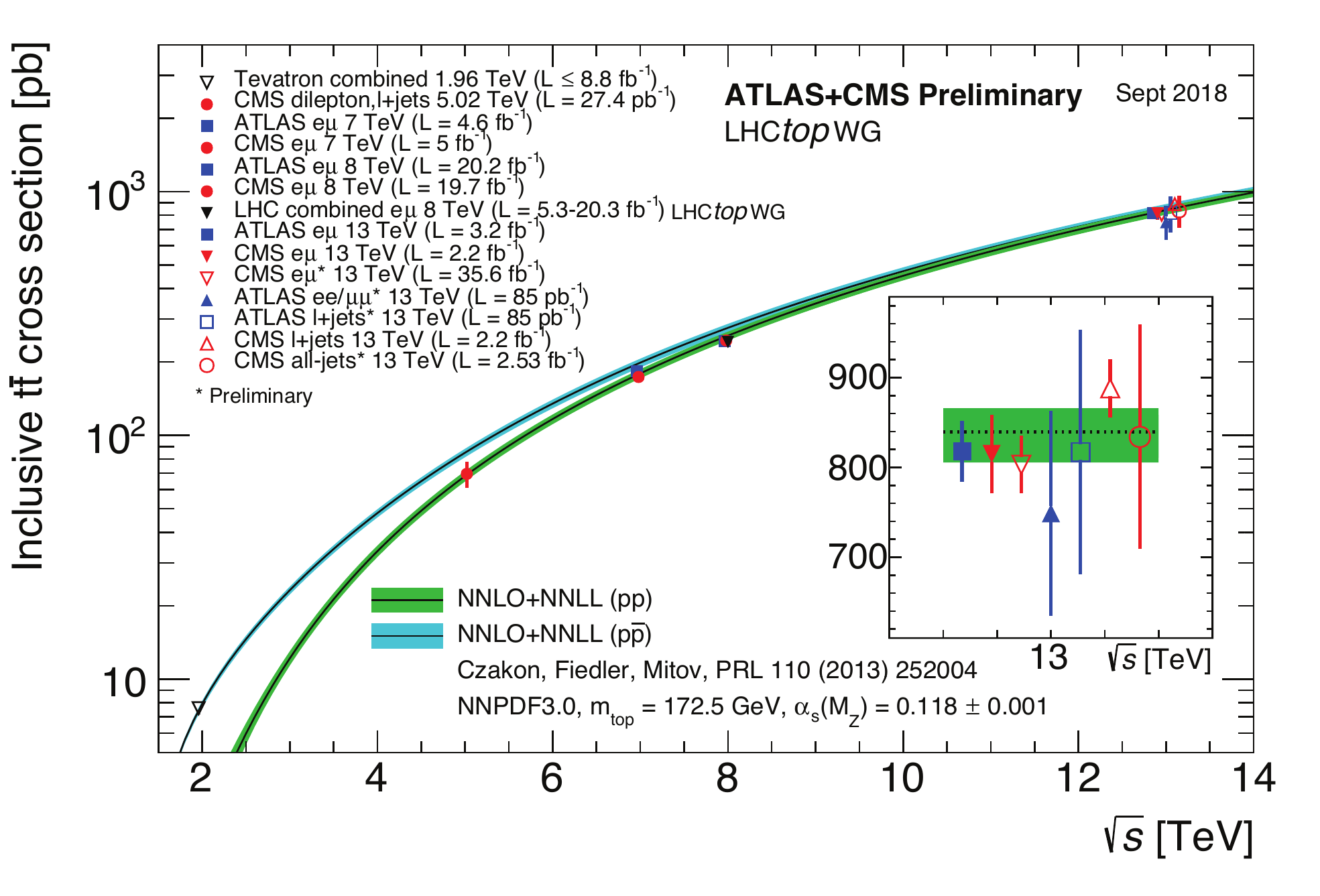}
\caption{Comparison of the measured values for the \ttbar production cross section as a function of the center of mass energy for \pbarp and pp collisions with NNLO QCD with NNLL resummation theoretical predictions.}  
\label{fig5}
\end{center}
\end{figure}

The large number of \ttbar events produced at LHC energies allows for full phase-space normalized differential cross section measurements as those in figure~\ref{fig6}, in which the ATLAS and CMS measurements are compared to theoretical predictions~\cite{TOPWG}. Overall good agreement is observed between data and predictions and the measurements can be used to improve the theoretical models.

\begin{figure}[ht]
\begin{center}
\includegraphics[width=7.5cm]{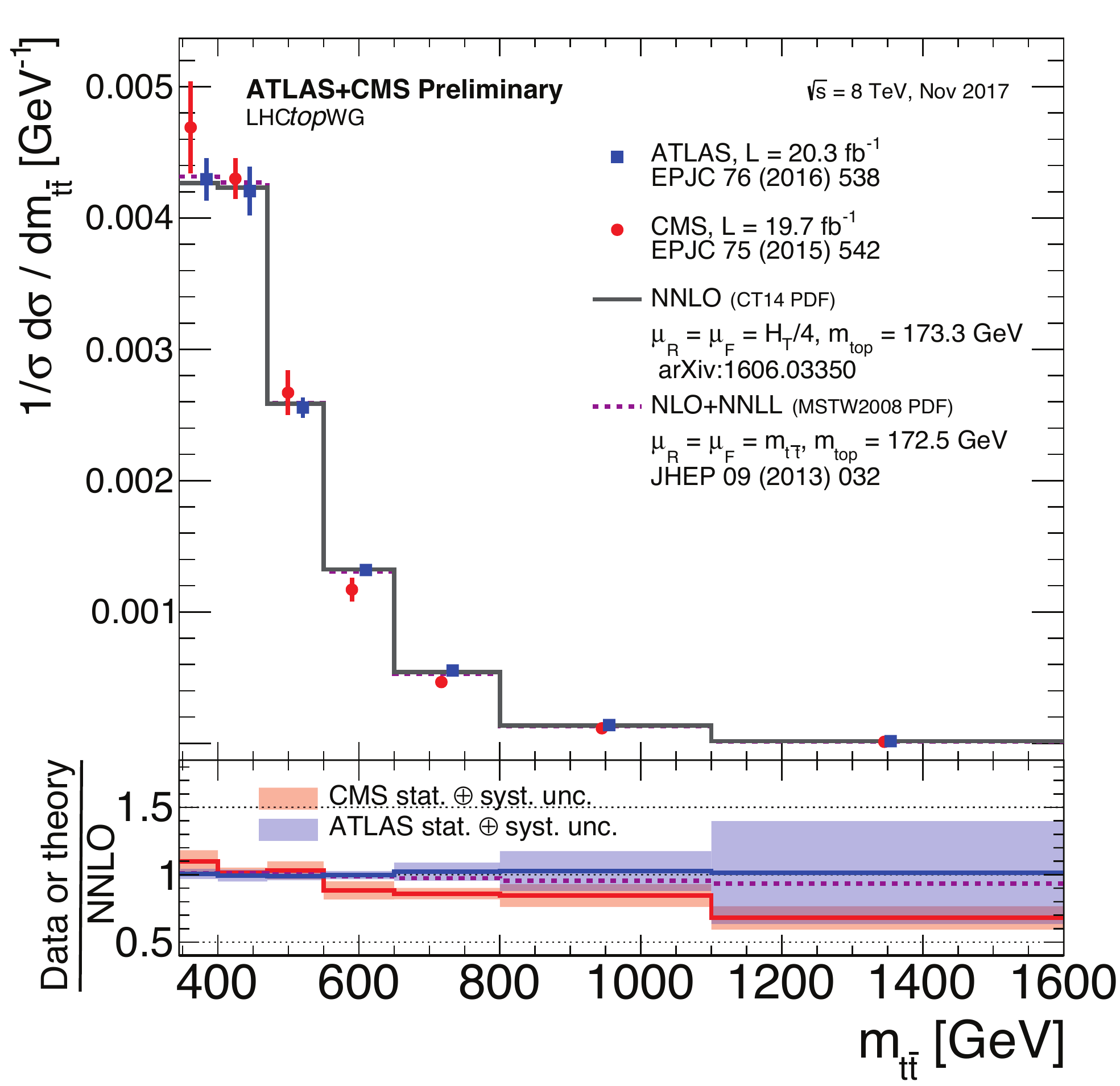}
\includegraphics[width=7.5cm]{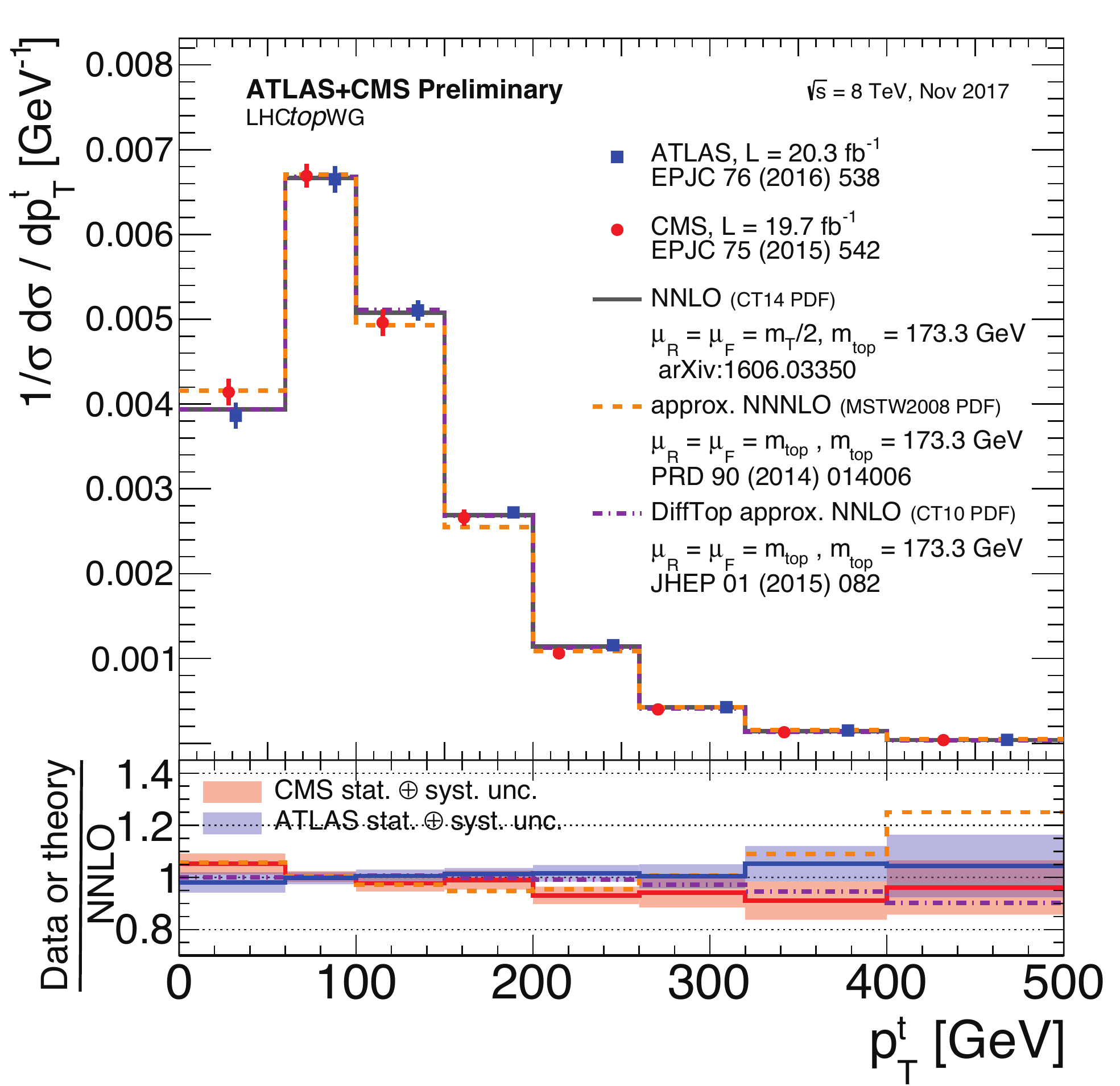}
\caption{Full phase-space normalized differential \ttbar cross-section as a function of the top quark pair invariant mass (left) and the transverse momentum of the top quark (right). The ATLAS and CMS results are compared to theoretical predictions at NNLO and NLO+NNLL.}
\label{fig6}
\end{center}
\end{figure}

Single top production at hadron colliders provides an 
opportunity to study the charged-current weak-interaction of the top quark. 
The SM predicts that the top quark decays almost exclusively to a 
$W$ boson and a bottom quark with $B(t \to Wb) \approx 1$. The rate for the process 
leads to a firm prediction for the top quark decay width $\Gamma_{t}$. 
A direct measurement of $\Gamma_{t}$ is of great importance, because the width would 
be affected by any non-expected decay modes of the top quark, whether they are 
observed or not. Unfortunately, $\Gamma_{t}$ cannot be directly measured in the 
\ttbar\ sample at hadron colliders, but its main component can be accessed 
through single top processes. If there are only three generations, the unitarity 
constraint of the CKM matrix implies that $|V_{tb}|$ is very close to unity. But, 
the presence of a heavy fourth generation quark with a large CKM coupling to the 
top quark could be consistent with large values of $B(t \to Wb)$, while resulting in 
an almost entirely unconstrained value for $|V_{tb}|$. A direct measurement of 
$|V_{tb}|$ can therefore explore the possibility of a fourth generation, and confirm 
that the top quark is indeed the SU(2) partner of the bottom quark. A measurement of the single top quark production 
cross section provides the only known way to directly measure $|V_{tb}|$ at a 
hadron collider. Figure~\ref{fig7} present the most recent ATLAS and CMS measurements of the single top production 
cross section in its three channels ($s$, $t$ and $Wt$ associated production) and for three center of mass energies, while figure~\ref{fig8} summarizes the corresponding extractions of the CKM matrix element $|V_{tb}|$~\cite{TOPWG}. Good overall agreement is observed between measurements and the predictions.

\begin{figure}[ht]
\begin{center}
\includegraphics[width=15cm]{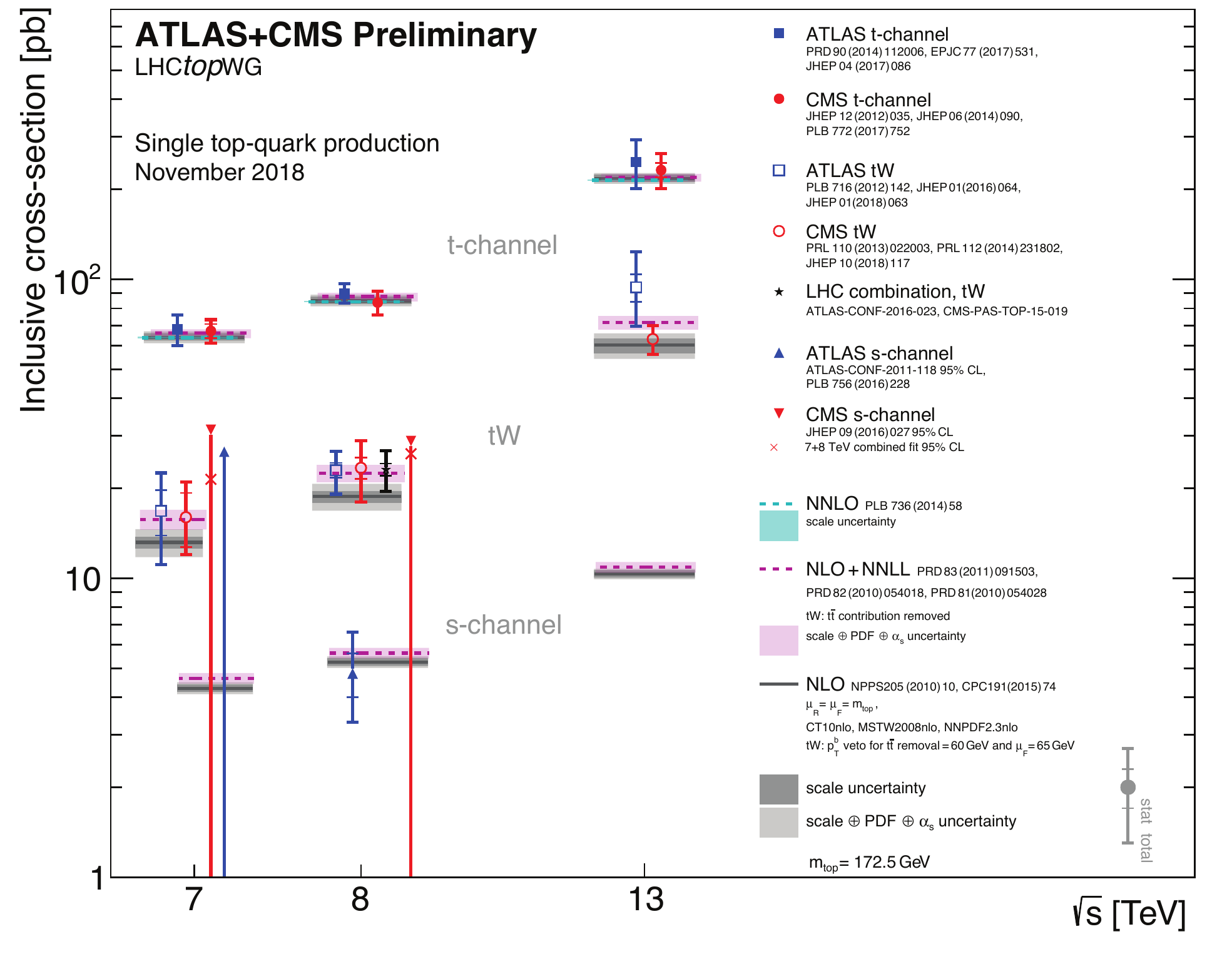}
\caption{Comparison of measured single top production cross sections as a function of the center of mass energy with various theoretical predictions. Good overall agreement is observed between measurements and the predictions.
}
\label{fig7}
\end{center}
\end{figure}

\begin{figure}[ht]
\begin{center}
\includegraphics[width=15cm]{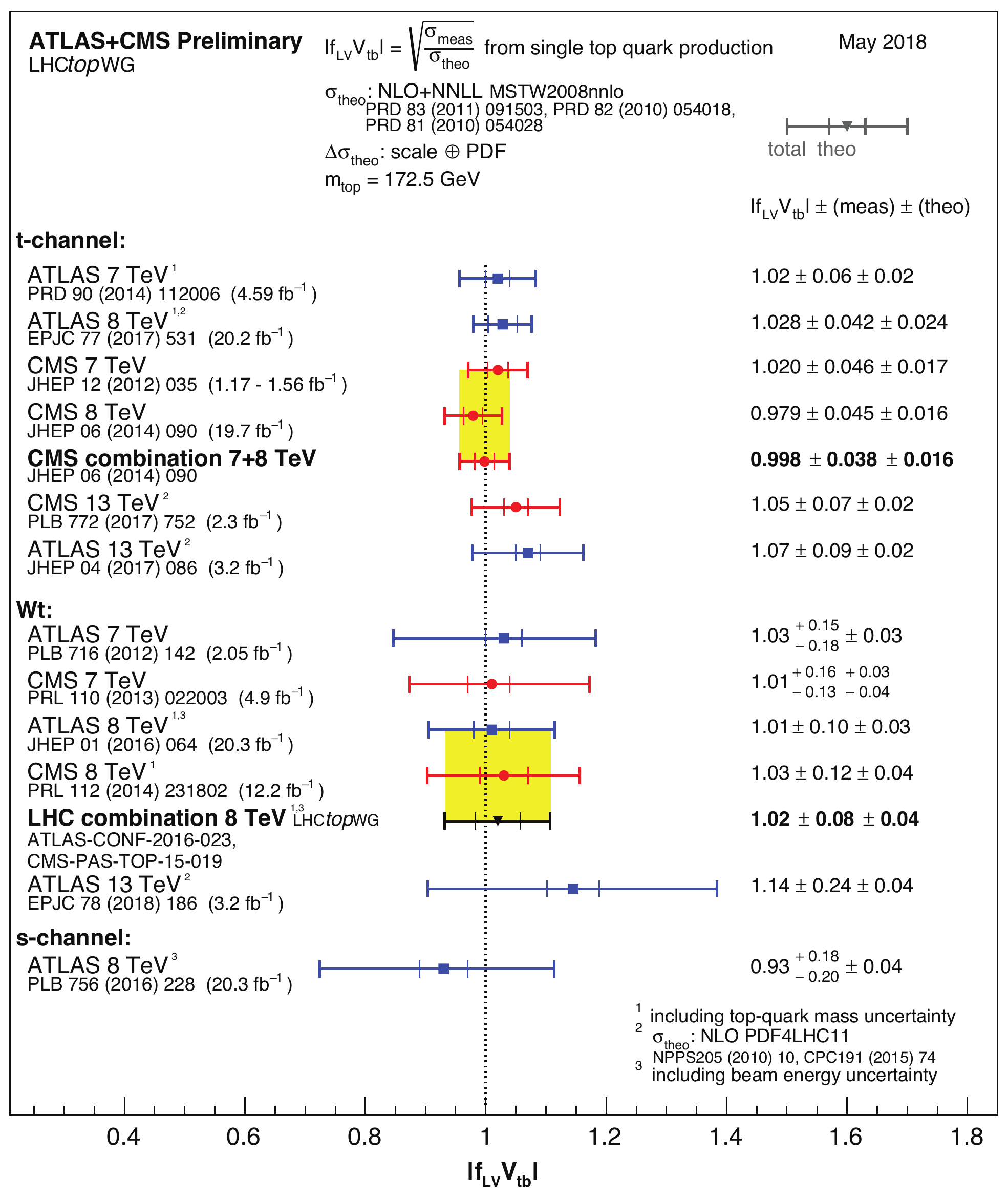}
\caption{Summary of the ATLAS and CMS extractions of the CKM matrix element Vtb from single top quark measurements. Good overall agreement is observed between measurements and the predictions.
}
\label{fig8}
\end{center}
\end{figure}

\clearpage
\section{Top Quark Properties}

Top quark properties are predicted by the SM and can be modified by BSM effects at the production or decay level. Precision measurements can uncover deviations of SM predictions which could serve as indirect evidence of BSM processes due to particles with masses not currently accessible at the LHC. Analyses typically rely on the \ttbar l+jets channel, due to its large samples, low background and constrained final state. Dedicated sensitive observables are defined for each property, with the LHC benefitting from and expanding the methods developed for those same studies at the Tevatron. 

From the many properties of interest for the top quark, few are so fundamental as its mass. The ATLAS and CMS collaboration have published measurements of the top quark mass using candidate events in all \ttbar decay channels. The top quark mass has also been measured from single top events and extracted indirectly from the \ttbar cross section measurement. Comparing precision measurements of the top quark mass, the $W$ and the Higgs boson masses with SM predictions provides a powerful tool to search for BSM effects. Deviations would be an indication of new physics in the mass loops. 

The most precise measurements of the top quark mass $m_t$ are obtained from the lepton+jets \ttbar sample, where the mass is reconstructed from a kinematic fit of the decay products to the \ttbar hypothesis. The mass is determined from a simultaneous fit of $m_t$ and the main uncertainties arising from the jet energy scale and the b-jet energy scale. The contributions from these uncertainties are statistical in nature, and will benefit from larger data samples. Figure~\ref{fig9} presents a summary of the direct ATLAS and CMS top $m_t$ measurements~\cite{TOPWG}, compared to the LHC and World averages. The results show good agreement between measurements and with the world average.

\begin{figure}[ht]
\begin{center}
\includegraphics[width=15cm]{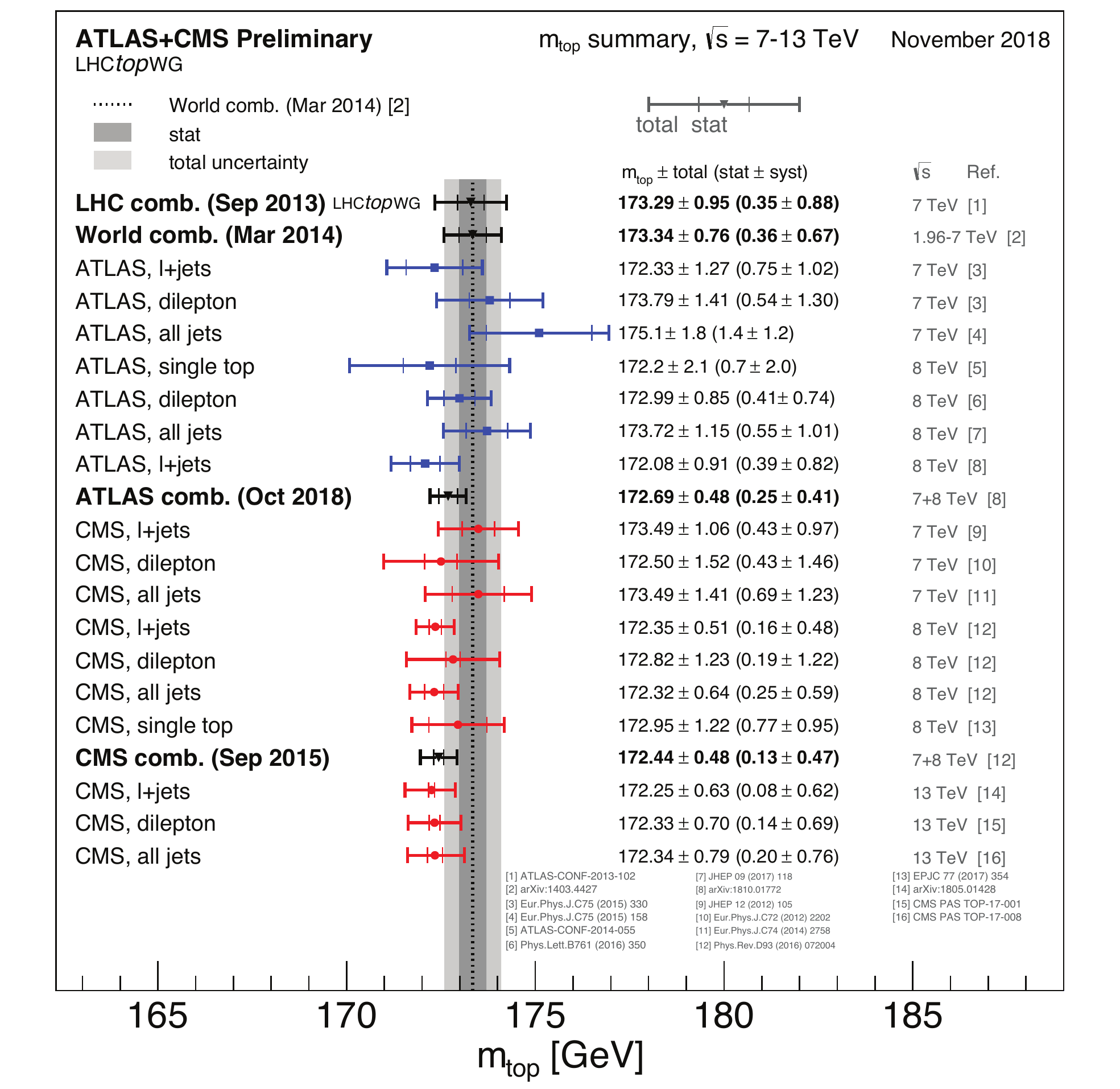}
\caption{Summary of the ATLAS and CMS direct top quark mass measurements compared to the LHC and Tevatron+LHC combinations. }
\label{fig9}
\end{center}
\end{figure}

A fundamental difference between the \ttbar production at the Tevatron and at the LHC is that the production is dominated by \qqbar annihilation in the former and by gluon fusion in the latter. At leading order, the SM predicts that the \ttbar production is forward-backward symmetric in \qqbar annihilation. However, higher order SM effects result in a small ($\approx 6.6\%$) positive asymmetry $A_{FB}$, such that the top (anti-top) quark is preferentially emitted in the direction of the incoming quark (anti-quark)~\cite{AFB}. BSM production mechanisms that exchange new bosons could enhance $A_{FB}$. There is no asymmetry in the gluon fusion \ttbar\ production that dominates at the LHC, but because quarks carry, on average, larger momentum than antiquarks, the rapidity distribution of top quarks is expected to be broader than that of anti-tops, which results in a \ttbar charge asymmetry $A_C$ of ($\approx 1\%$). Early Tevatron $A_{FB}$ measurements~\cite{cdf,d0}, based on about half the data that would eventually become available, sparked a huge interest when they showed larger asymmetries than those predicted by the SM at the time~\cite{sm}, especially because the discrepancies grew with larger top quark pair masses and rapidity difference. Measurements using the full Tevatron dataset, and combining the results from the two collaborations, recently became available~\cite{cdfd0}. Even though all measurements favor somewhat larger positive asymmetries than the predictions~\cite{sm2}, none of the observed differences are larger than 2 standard deviations, as can be seen in figure~\ref{fig10}. The ATLAS and CMS collaborations have combined their inclusive and differential measurements of $A_C$ at two center of mass energies (7 and 8~TeV), obtaining $A_C = 0.005 \pm 0.007 {\rm(stat)} \pm 0.006 {\rm (syst)}$ and  and $A_C = 0.0055 \pm 0.0023 ({\rm(stat)} \pm 0.0025 {\rm (syst)}$ at 7 and 8~TeV, respectively, in good agreement with the respective SM predictions~\cite{atlascms}. Figure~\ref{fig11} shows the measured combined inclusive $A_C$ at 8~TeV versus the combined Tevatron $A_{FB}$ compared with the SM prediction at NNLO+EW NLO~\cite{sm3} and various BSM predictions that could affect the asymmetries. The combined Tevatron/LHC measurements uniquely restrict the phase space of possible BSM phenomena which would produce large asymmetries, including models that predict the existence of heavy $W'$ bosons, heavy axigluons, scalar isodoublets, color triplet scalars and color sextet scalars~\cite{bsm}. 

\begin{figure}[ht]
\begin{center}
\includegraphics[width=12cm]{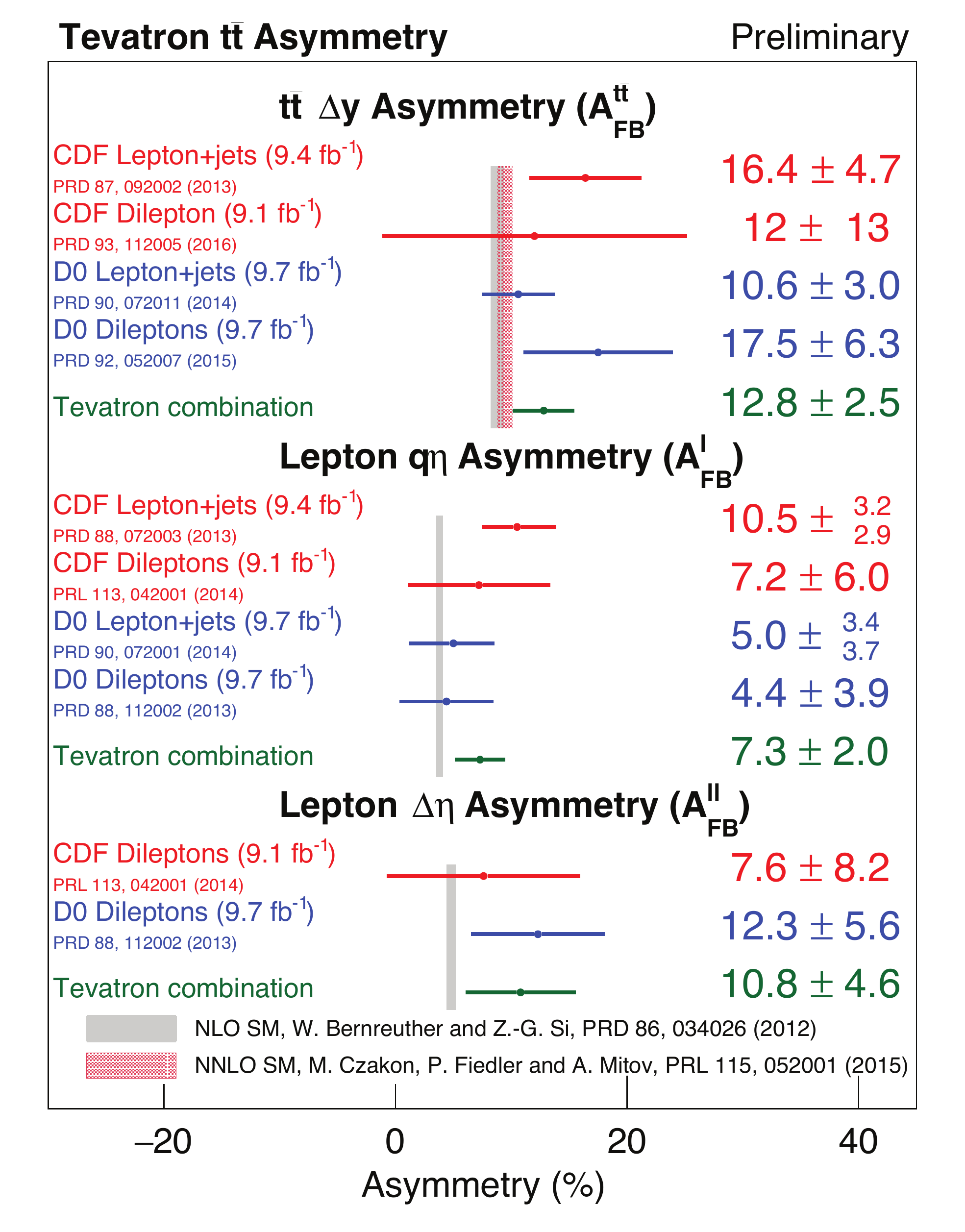}
\caption{Summary of inclusive forward-backward asymmetries in \ttbar events at the Tevatron. Even though all measurements favor somewhat larger positive asymmetries than the predictions~\cite{sm2}, none of the observed differences are larger than 2 standard deviations.}
\label{fig10}
\end{center}
\end{figure}

\begin{figure}[ht]
\begin{center}
\includegraphics[width=12cm]{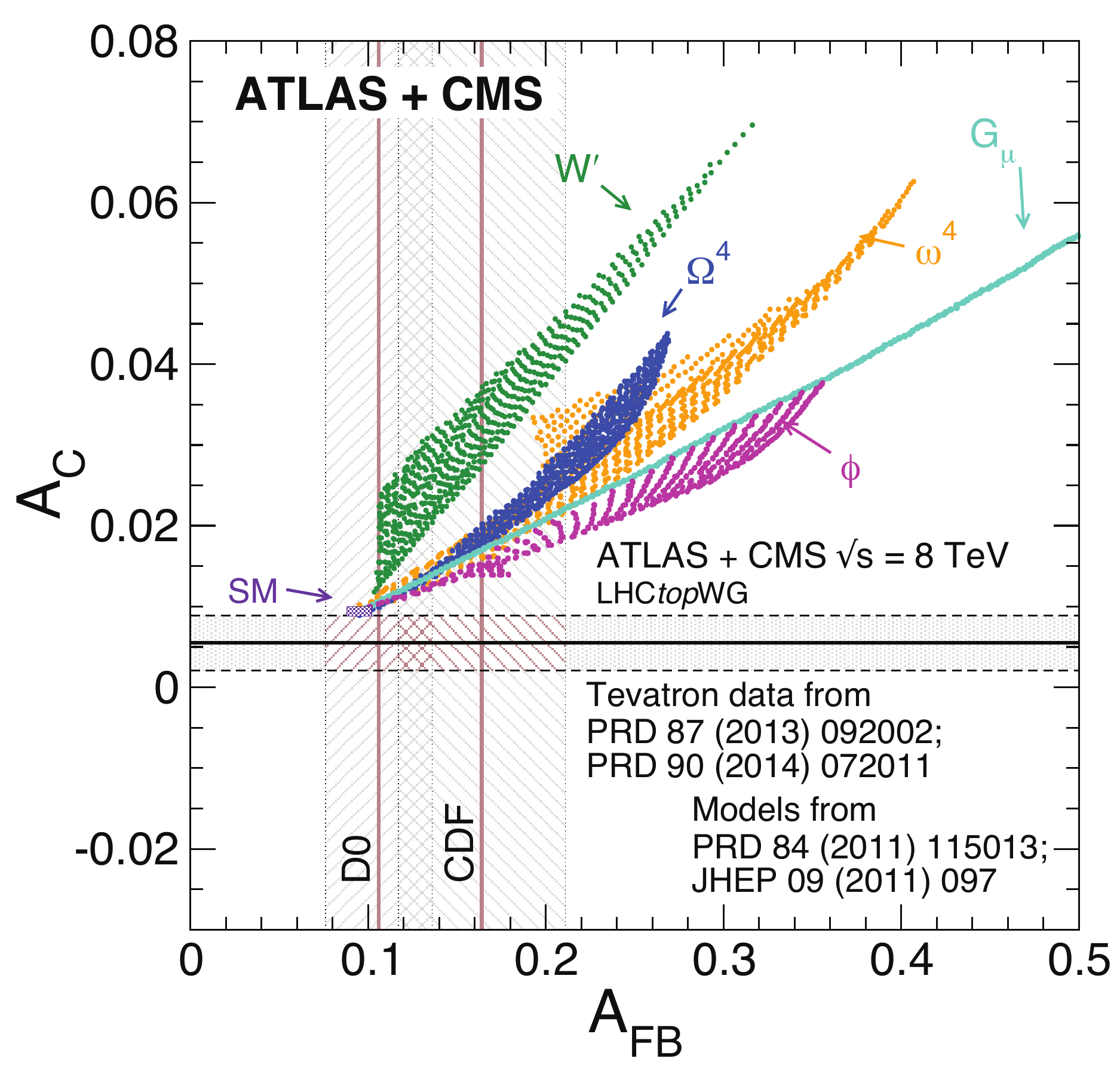}
\caption{Combined inclusive LHC $A_C$ at 8~TeV versus the combined Tevatron $A_{FB}$ compared with the SM and BSM predictions. The combined Tevatron/LHC measurements uniquely restrict the phase space of possible BSM phenomena which would produce large asymmetries.}
\label{fig11}
\end{center}
\end{figure}

A property characterizing the dynamics of the top-quark decay is the helicity state of the on-shell $W$ boson. The $W$ boson can have three possible helicity states, and the fractions of $W^+$ bosons produced in these states are denoted as $f_0$ (longitudinal), $f_-$ (left-handed), and $f_+$ (right-handed). In the SM, the top quark decays through the $V-A$ weak charged-current interaction, which strongly suppresses right-handed $W^+$ bosons or left-handed $W^-$ bosons.   Significant deviations from these expectations would indicate either a departure from the $V-A$ structure of the $tWb$ vertex or the presence of a non-SM contribution to the $t\bar t$ candidate sample.
Both ATLAS and CMS have measured the helicity fractions to a precision better than $5\%$~\cite{TOPWG} and found good agreement with the SM prediction at NNLO~\cite{sm4}, as shown in figure~\ref{fig12}. 

\begin{figure}[ht]
\begin{center}
\includegraphics[width=15cm]{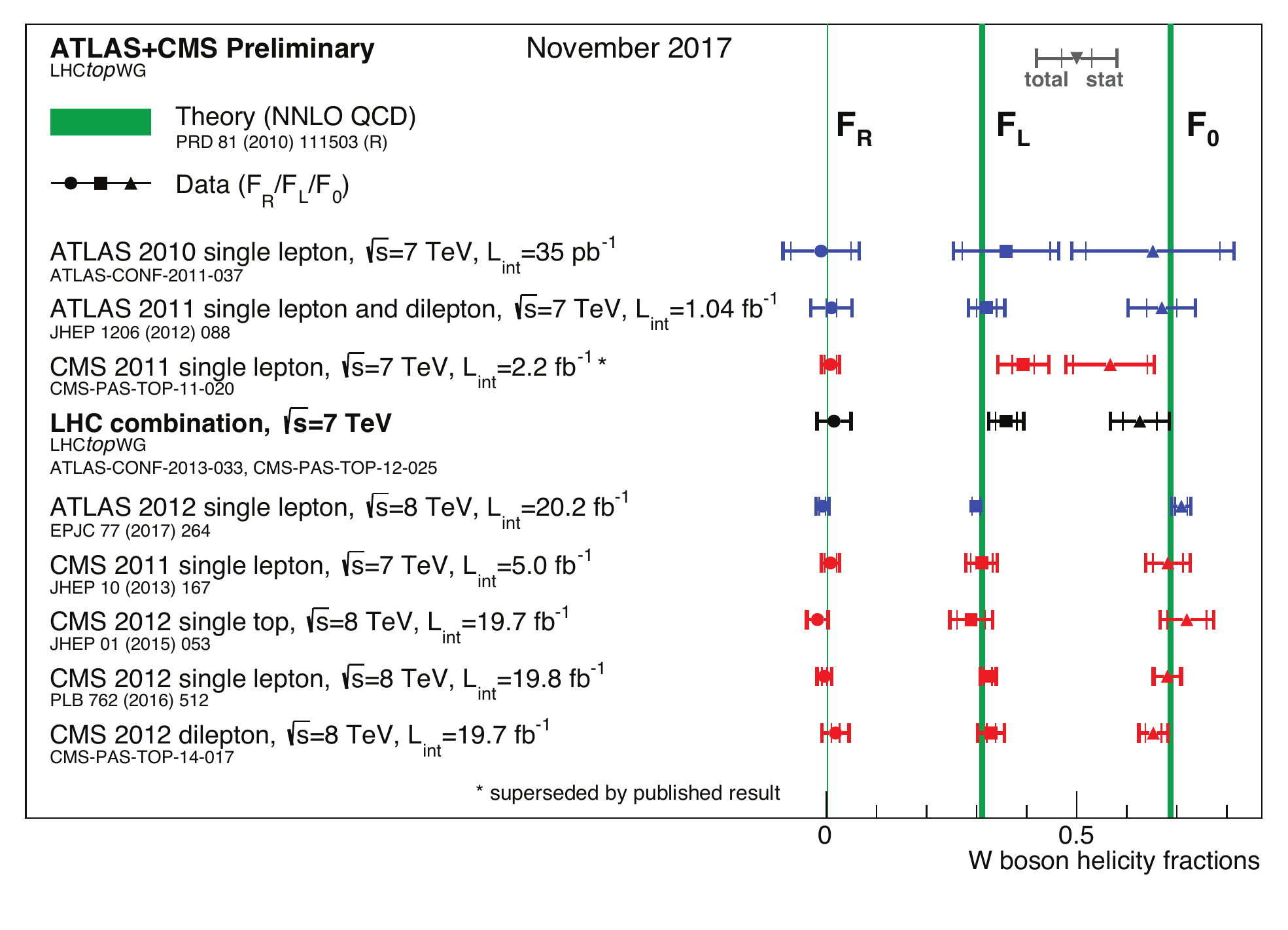}
\caption{Summary of the $W$ boson helicity fractions from top quark decays measured in \ttbar events collected at 7 and 8~TeV by the ATLAS and CMS collaboration. The results, which have a precision better than $5\%$, agree well with SM NNLO predictions.}
\label{fig12}
\end{center}
\end{figure}

\clearpage
\section{Boosted Top Quark Production}
At Tevatron energies, the majority of \ttbar events were produced at rest. This has changed at the LHC, where the higher center of mass energies result in top quarks that are often produced with a high Lorentz-boost in momentum, which yields decay products that are partially or fully merged because the angular distance between partons is smaller than the jet clustering distance parameter. As a consequence, the three quarks from the hadronically decaying top quark may be reconstructed as one fat jet, and similarly, for the leptonically decaying top quark, the lepton may appear as non-isolated due to its proximity to the b-quark. Special techniques were developed by both ATLAS and CMS to reconstruct these boosted top quarks, in both the hadronic and the leptonic case. The cornerstone of these techniques relies on  the ability to 
reconstruct a single jet that contains the full energy of the decay. This ``fat jet'' is then distinguished from ordinary jets through the identification of the jet internal structure. In the case of the hadronic decay of the top quark, the internal structure can identify the b-jet and the individual light jets that result from the $W$ boson decay. In the case of the leptonic decay of the top quark, the energy deposits from the leptons are identified and excluded from the b-jet reconstruction. This is particularly challenging in the case of the electron, as no isolation requirement can be applied. 

The ATLAS collaboration has measured the \ttbar differential production cross section at $\sqrt{s} = 8\;\rm {TeV}$ for lepton+jets events with high transverse momentum. The measurement is reported as a function of the transverse momentum of the hadronically decaying top quark for values of $p_T > 300\;\rm GeV$~\cite{ATLAS-boosted}. Figure~\ref{fig13}~(left) shows the $p_T$ of the leading fat jet compared to the SM prediction. The lower panel shows the ratio of the MC prediction to the data, where good agreement is observed within the uncertainties. 

The CMS collaboration has searched for boosted anomalous resonant \ttbar production in events with zero, one and two leptons~\cite{CMS-boosted}, by reconstructing the \ttbar invariant mass distribution and focusing in the area of masses of at least 1~TeV. Figure~\ref{fig13}~(right) shows the \ttbar invariant mass for events in the muon+jets channel in the cases in which the hadronic top has been reconstructed as a fat jet. No excess of data over the SM predictions is observed, and limits are set on the production cross section times branching fraction, probing a region of parameter space for certain models of new physics not yet constrained by precision measurements. 

\begin{figure}[ht]
\begin{center}
\includegraphics[width=7.5cm]{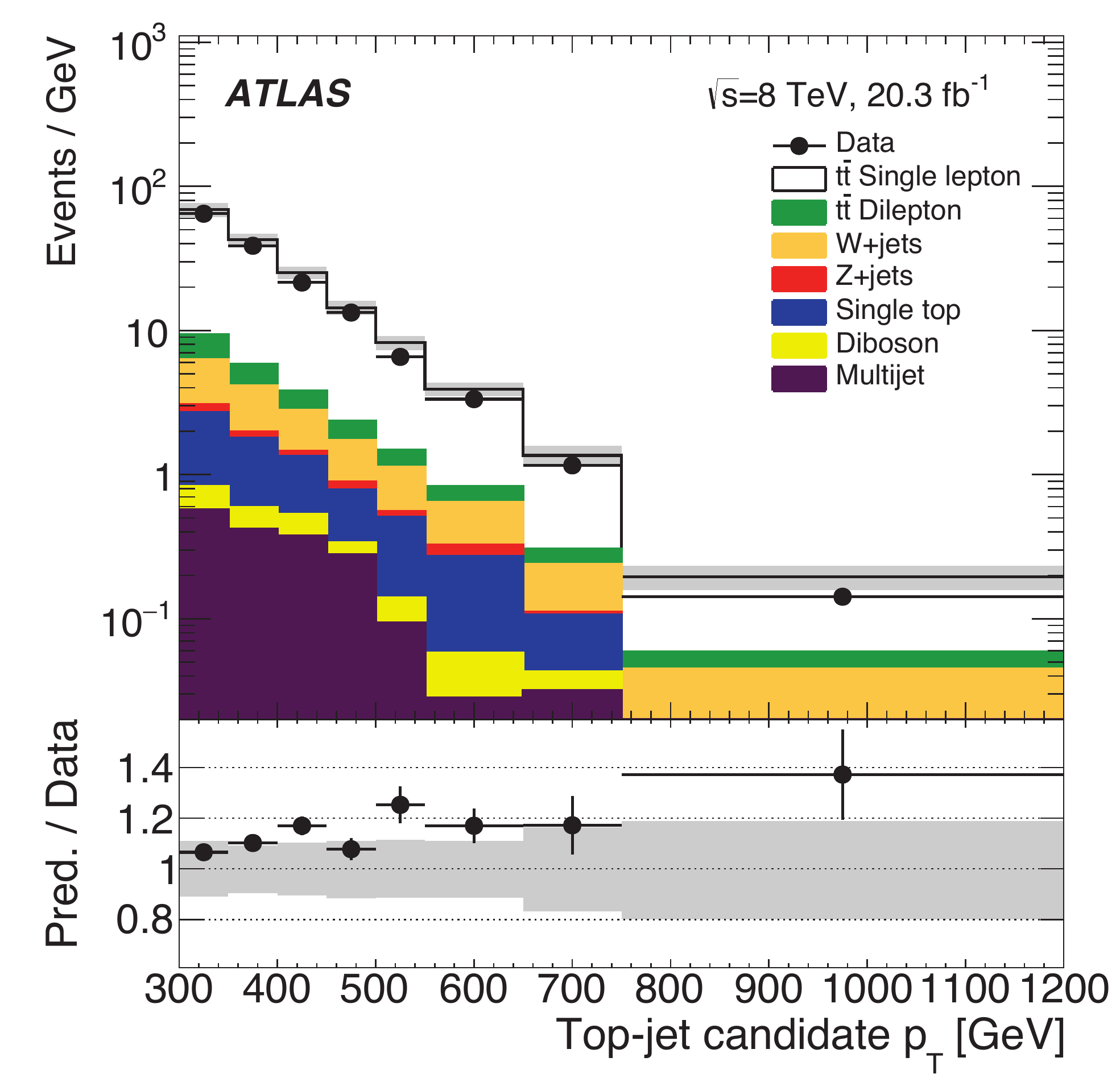}
\includegraphics[width=8cm]{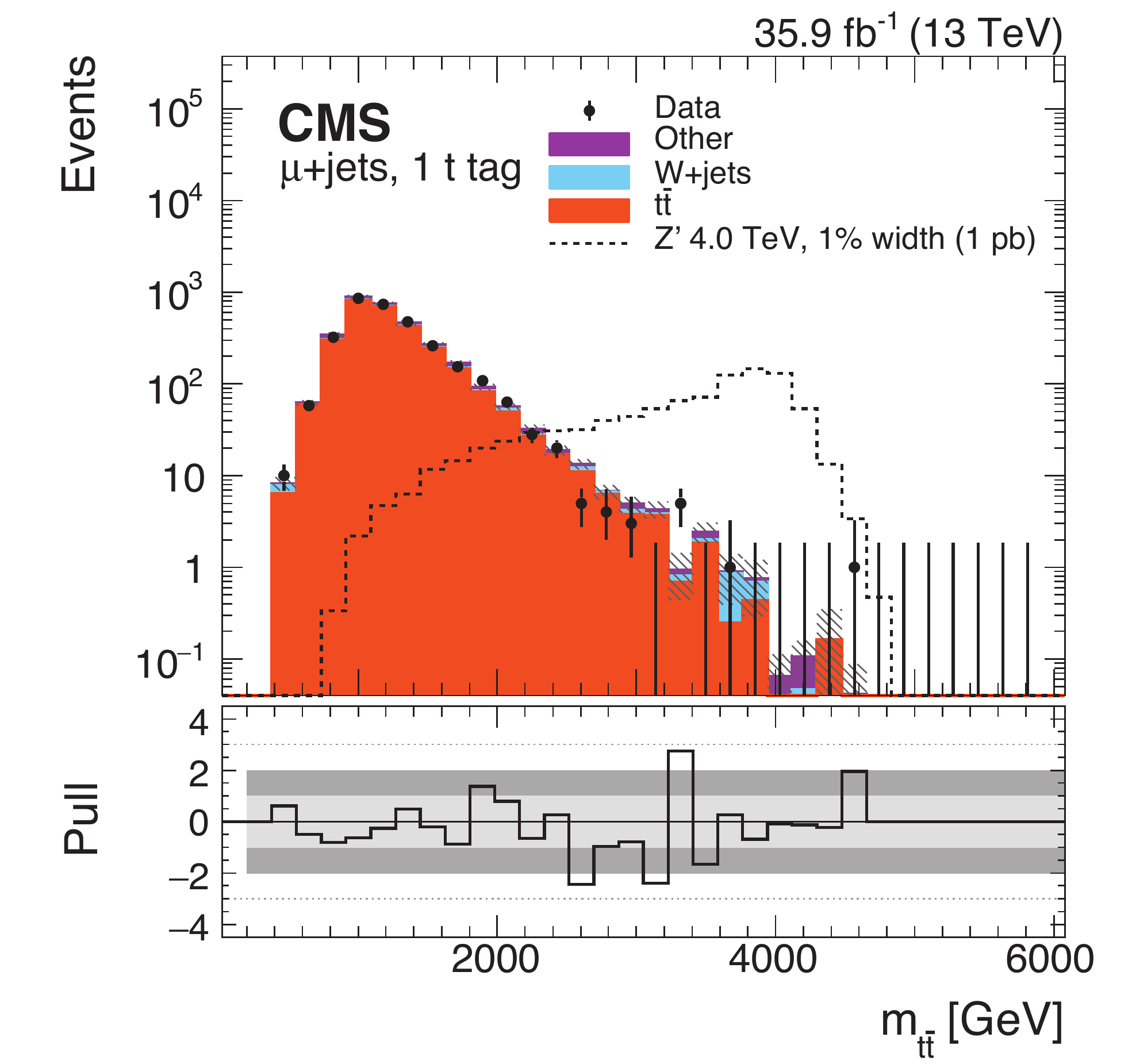}
\caption{\ttbar differential production cross section in the lepton+jets channels as a function of the $p_T$ of the hadronically decaying top quark for values of $p_T > 300\;\rm GeV$ (left). Invariant mass of \ttbar events in the boosted lepton+jets channel  in the cases in which the hadronic top has been reconstructed as a fat jet (right). 
The lower panels show the ratio of the MC prediction to the data, where good agreement is observed within the uncertainties.}
\label{fig13}
\end{center}
\end{figure}

\clearpage
\section{Higgs Boson Studies}
The Higgs Boson, that had been predicted by the SM since the seventies, was observed on July 4, 2012 by the ATLAS~\cite{ATLAS-higgs} and CMS~\cite{CMS-higgs} collaborations. The analyses used the fully reconstructed Higgs boson decay products and their excellent mass resolution to drive the discovery; however, the channel in which the Higgs boson decays into a \bbbar pair, even though it represents 58\% of the branching ratio, was not observed.  In the years since the discovery, both collaborations worked hard  to extract the $H\to b\overline{b}$ signal from the overwhelming multijet \bbbar background taking advantage of the associated production, in which the Higgs boson is produced together with a $W$ or a $Z$ boson. That process had already been used by the LEP collaborations to exclude Higgs masses below $114.4\;\rm GeV$~\cite{LEP-higgs}, and by the Tevatron collaborations to claim evidence of a observation of the Higgs boson with a significance of $3.3\sigma$~\cite{Tevatron-higgs}. At the LHC, both collaborations select events with zero, one or two charged leptons (electrons or muons) to include the cases in which the $Z$ boson decays to two neutrinos, and the leptonic decays of the $W$ and the $Z$ boson, respectively. In all cases, the events are required to include two b-jets. Several tools were developed to make the signal accessible. First, the vector boson was required to be highly boosted, which had the effect of suppressing the QCD multijet background and also of merging the \bbbar pair so that it was reconstructed as a single fat jet. Jet substructure techniques and a special b-jet tagger based on a deep neural network discriminant were developed to identify the \bbbar jet with high efficiency and a low 0.1\% misidentification rate. The resolution of the \bbbar invariant mass was improved by applying a multivariate regression technique. A deep neural network was then used together with 7 signal and 21 background regions to simultaneously extract the background normalization and the signal. Figure~\ref{fig14} shows the output of the multivariate discriminant and the dijet invariant mass distribution. Both collaborations obtained significances larger than $5\sigma$ and the observation of the $VH(b\overline{b})$ process~\cite{ATLAS-hbb,CMS-hbb}. Both collaborations also measured the signal strength, defined as the ratio of the number of observed $H\to b\overline{b}$ events over the number predicted by the SM, to be consistent with 1 within an uncertainty of about $20\%$. 

Seven years after the Higgs boson observation, Higgs physics has entered a precision era. The $\gamma \gamma$, $ZZ$ and $WW$ decays were firmly established and the Higgs mass has been measured to a $0.15\%$ precision. The Yukawa mechanism has been established in the last 2 years by the observation of the $\tau \tau$ and \bbbar decays and the ttH process. And differential cross section measurements are being used to compare the data to state-of-the-art calculations. But there is still a lot to learn, in particular, searching for di-Higgs production is vitally important to start to understand the self-couplings of the Higgs.

\begin{figure}[ht]
\begin{center}
\includegraphics[width=7.5cm]{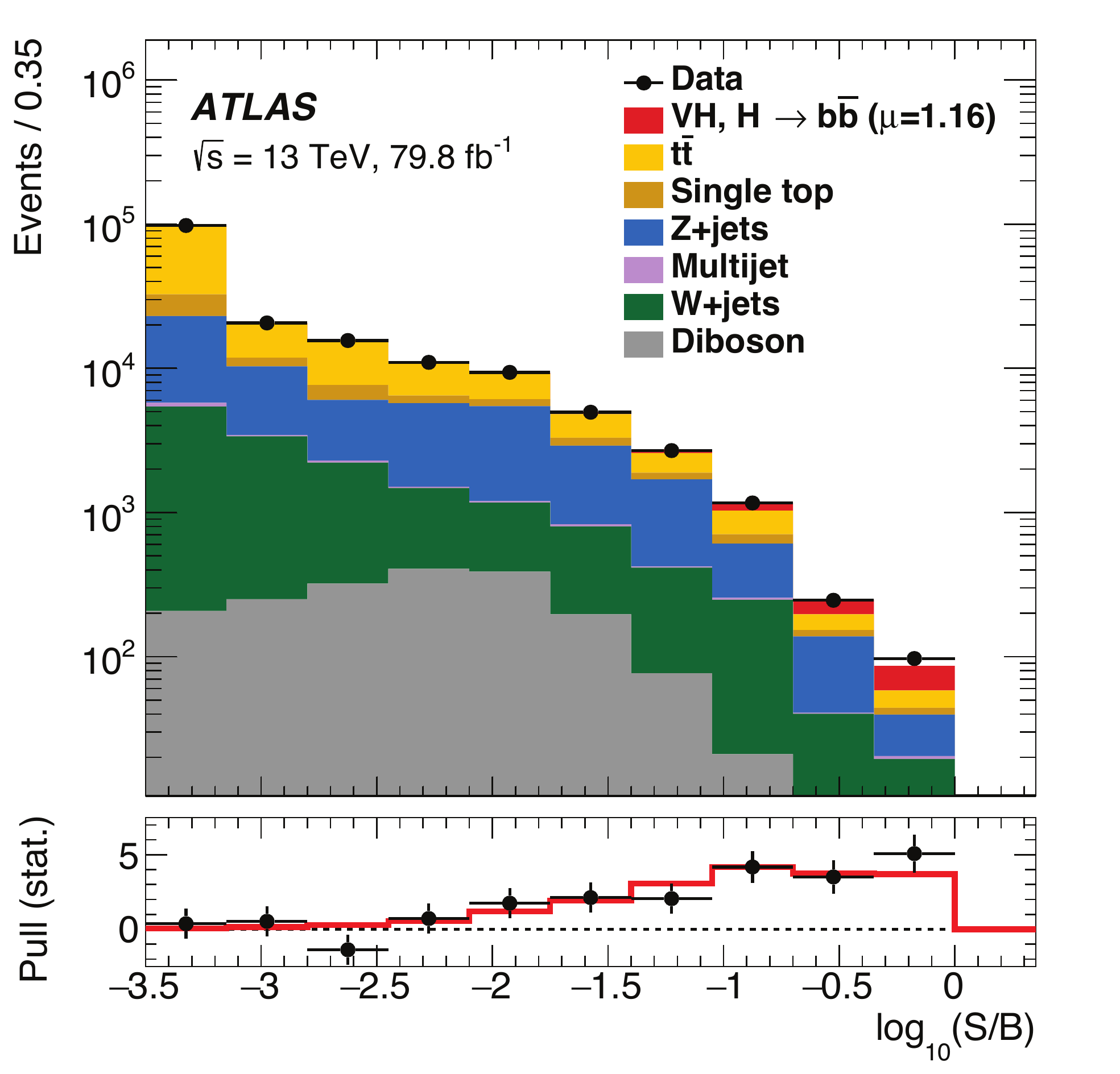}
\includegraphics[width=7.5cm]{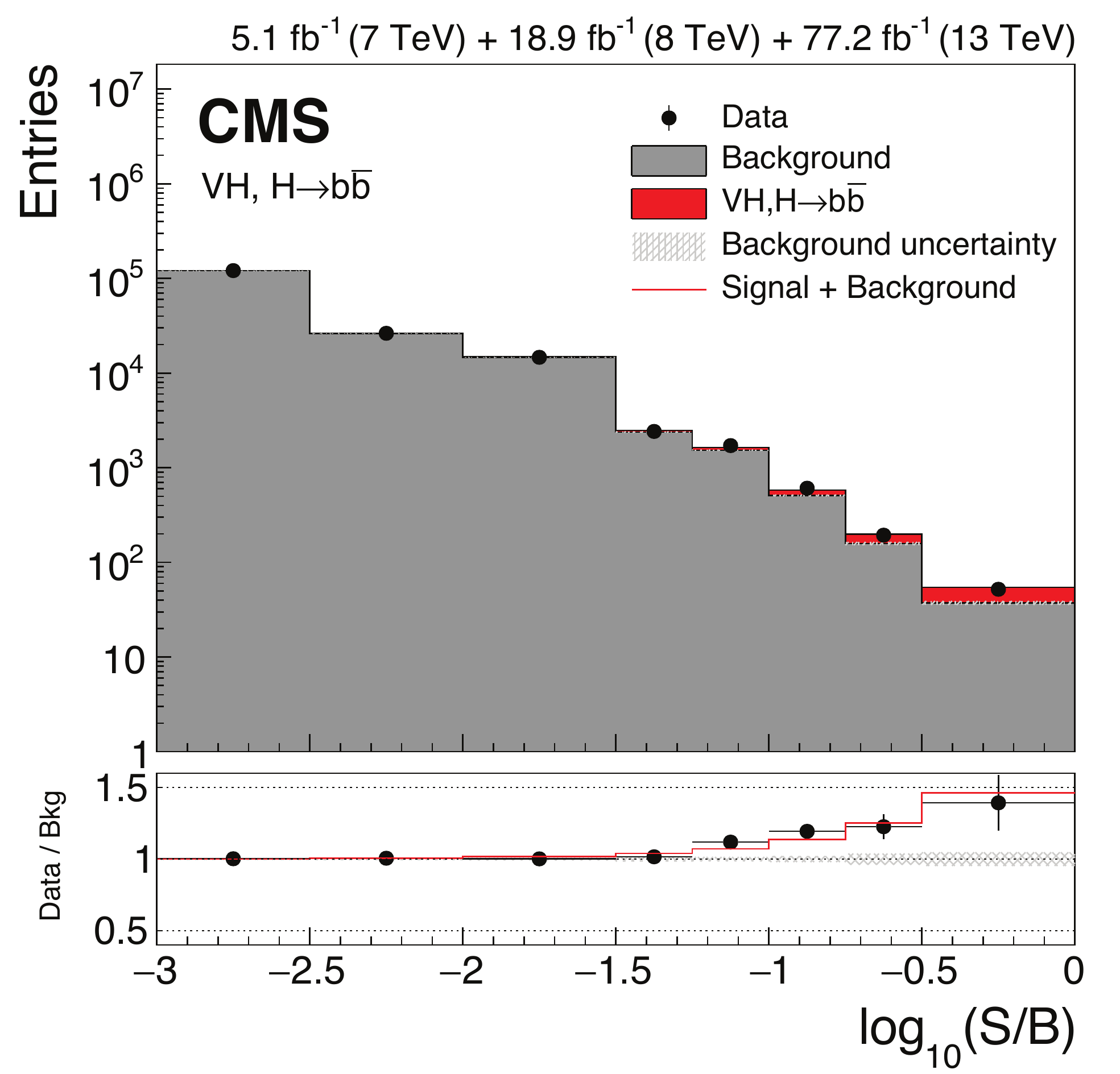}
\includegraphics[width=7.5cm]{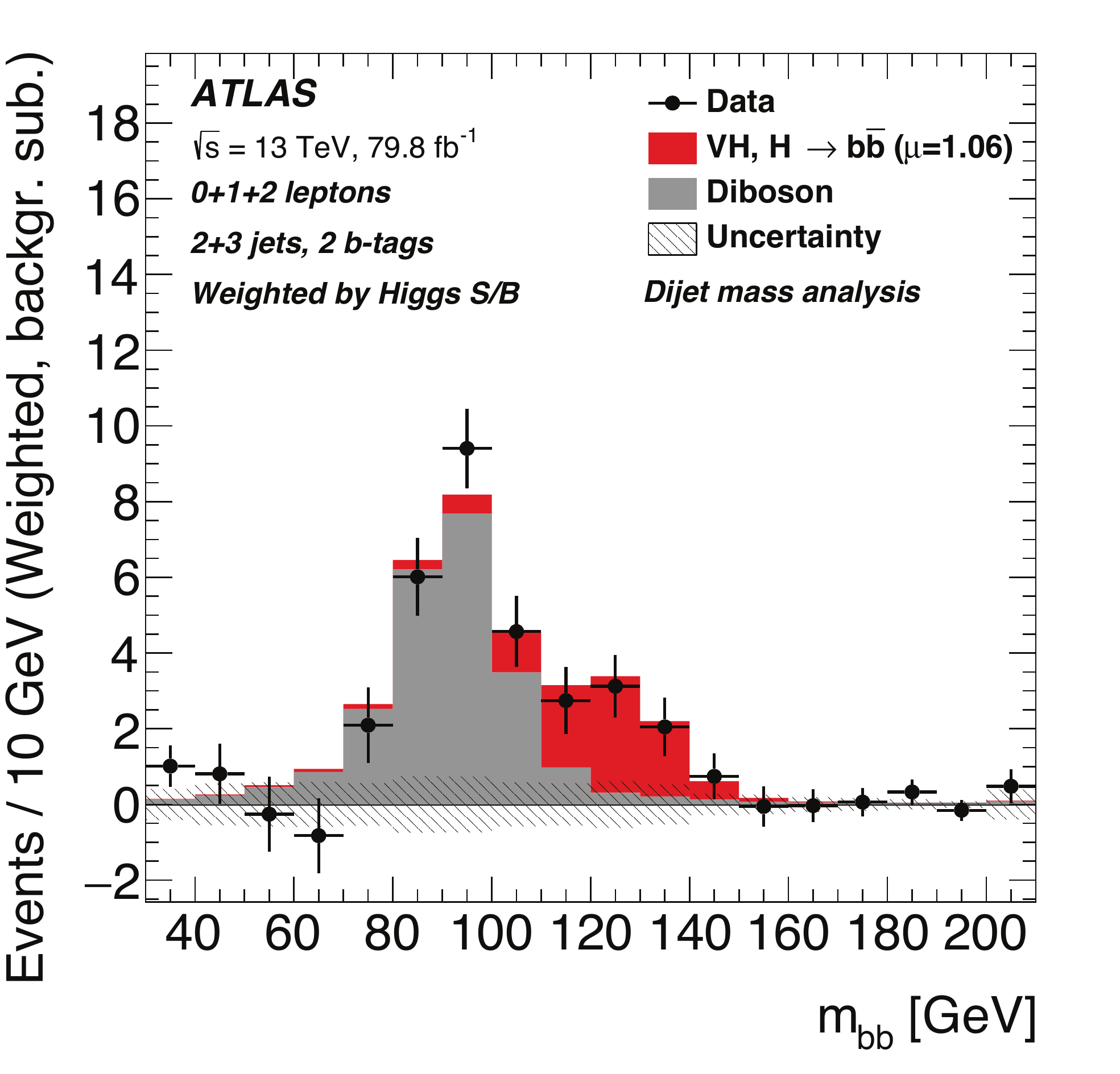}
\includegraphics[width=7.5cm]{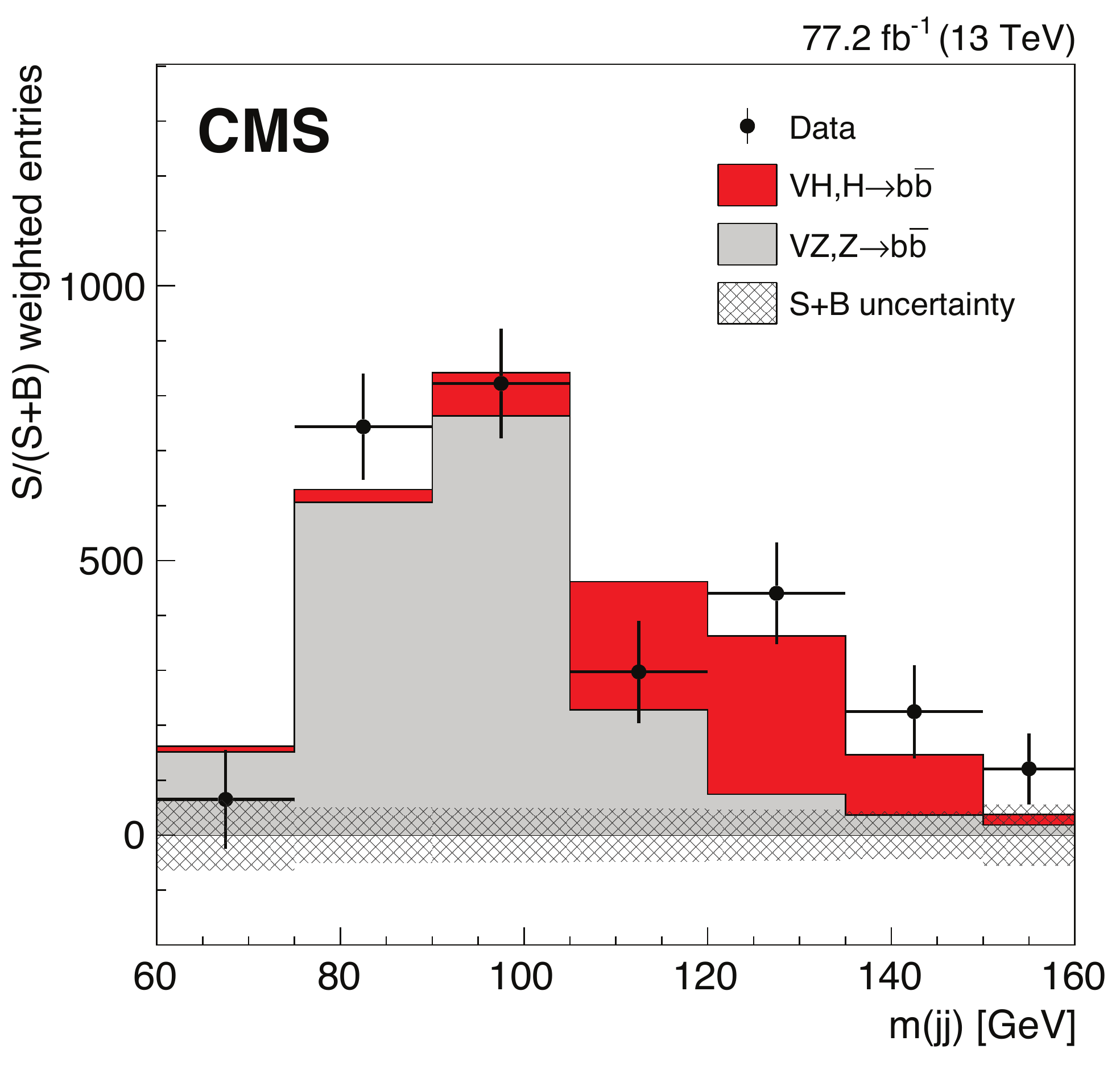}

\caption{Top: Event yields as a function of the logarithm of signal over background for data, background and a $125\;\rm GeV$ Higgs boson for ATLAS (left) and CMS (right). Bottom: The distribution of the invariant mass of the \bbbar pair, with the Higgs boson signal shown in red.}
\label{fig14}
\end{center}
\end{figure}

\clearpage
\section{Searches for Beyond the Standard Model Processes}
Since its inception more than 50 years ago, the SM has been extremely successful in predicting the existence of particles and processes that were later observed in experiments. In particular, the ATLAS and CMS collaborations have developed a broad and rich program, probing processes that span 9 orders of magnitude in production cross section. Some of the rare processes had not been observed before and are being produced at rates comparable with those expected from BSM predictions.  Figure~\ref{fig15} shows a summary of SM total production cross section measurements, corrected for leptonic branching fractions, compared to the corresponding theoretical expectations and ratio with respect to best theory. Figure~\ref{fig16} shows a summary of the cross section measurements for SM processes by the CMS collaboration.

\begin{figure}[ht]
\begin{center}
\includegraphics[width=15cm]{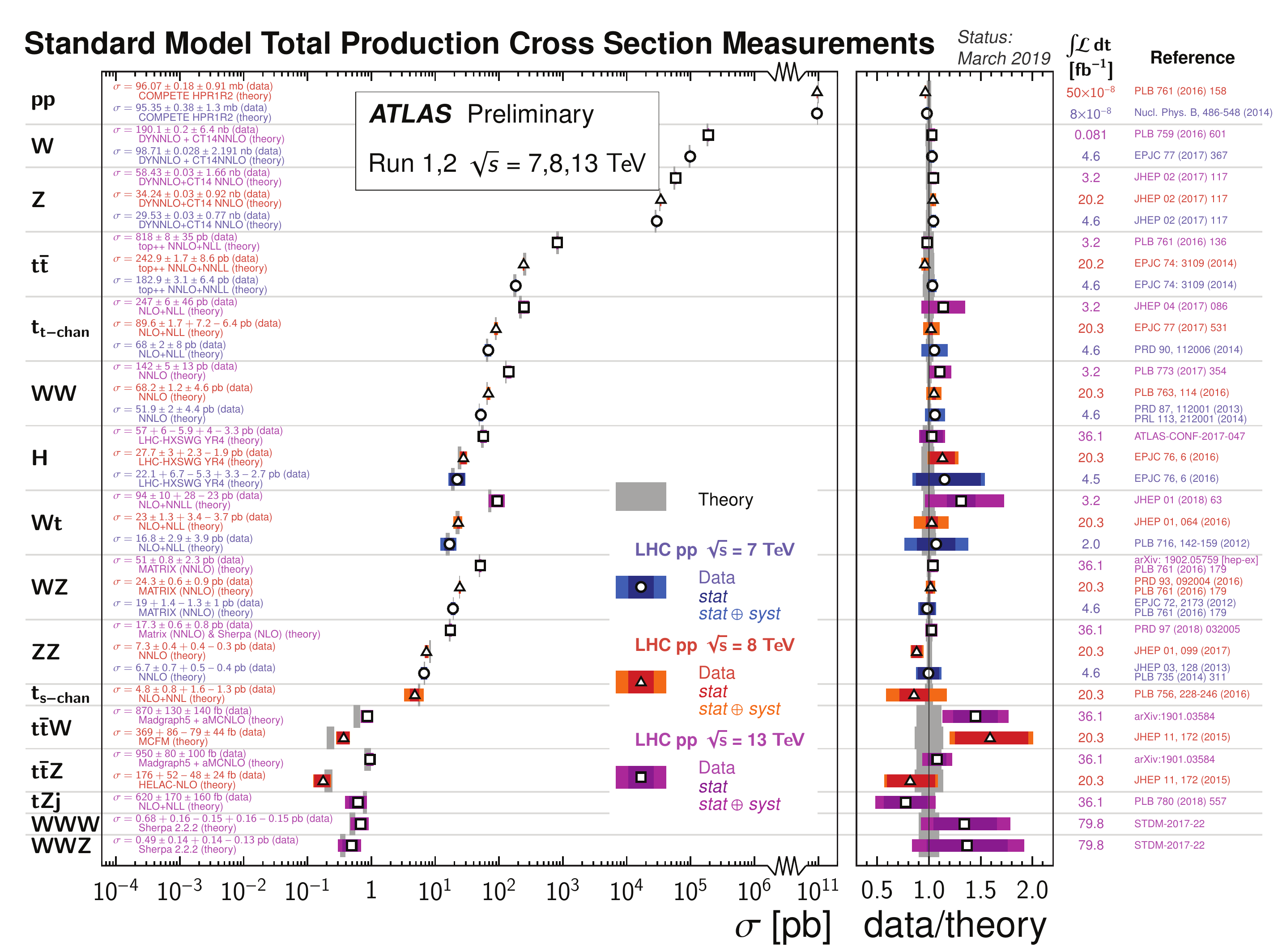}
\caption{Summary of measured SM production cross sections reported by the ATLAS collaboration, corrected for leptonic branching fractions, and compared to the corresponding theoretical expectations.}
\label{fig15}
\end{center}
\end{figure}

\begin{figure}[ht]
\begin{center}
\includegraphics[width=15cm]{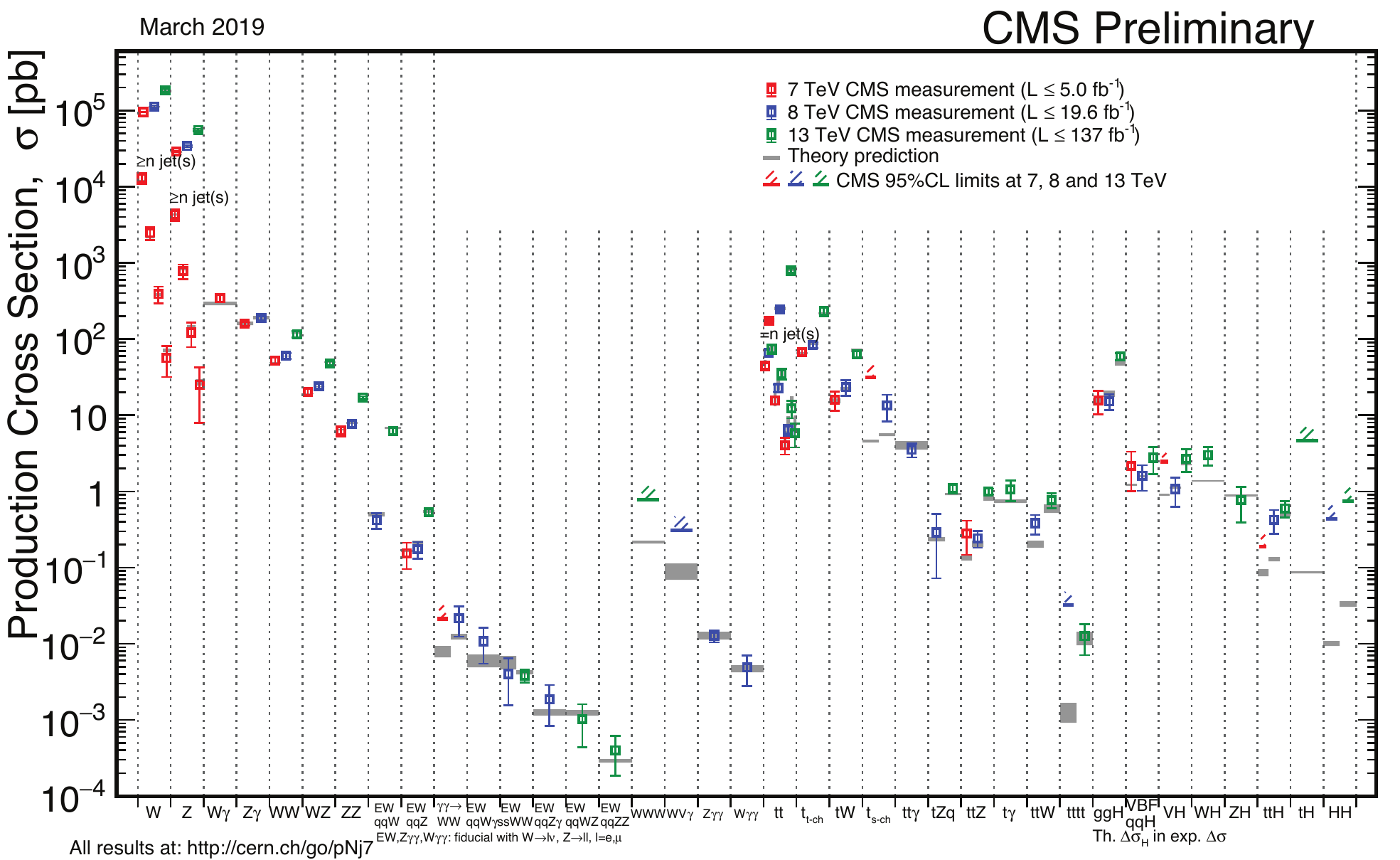}
\caption{Summary of the measured cross sections for SM processes as reported by the CMS collaboration. Each result is compared to the best available theoretical prediction.}
\label{fig16}
\end{center}
\end{figure}

In spite of the SM success, there remain several unanswered questions: why are there exactly three generations of quarks and leptons?
Are quarks and leptons actually fundamental, or made up of even more fundamental particles? Why can't the SM predict a particle's mass? How come neutrinos have mass?
Why do we observe matter and almost no antimatter if we believe there is a symmetry between the two in the universe? What is this "dark matter" that we can't see that has visible gravitational effects in the cosmos? How does gravity fit into all of this? And the uncomfortable issue of fine tuning. The BSM front runner around 2007 was SuperSymmetry (SUSY), that hypothesizes that a symmetry exists between fermions and bosons in which each boson has a fermion super-partner with the same mass and quantum numbers and vice-versa. These superpartners contribute with opposite sign to the loop corrections to the Higgs mass providing cancellation of the divergent terms. SUSY is theoretically compelling, providing a solution to the Higgs hierarchy problem, allowing unification of gauge couplings, and even predicting a dark matter particle candidate. For experimentalists SUSY predicted the existence of a large number of new particles, giving rise to striking experimental signatures ready to be discovered. The LHC years have witnessed a systematic exploration of TeV scale gluinos and squarks, and the LHC collaborations are just starting to gain sensitivity to Higgsinos, which in many models are expected to be the lowest mass SUSY particles with masses around a few hundreds GeV. Both collaborations are also in the process of extending their searches to 
unexplored regions of parameter space characterized by challenging manifestations of SUSY. In parallel to the continuing exploration of the SUSY parameter space, both collaborations have also vigorously pursued a plethora of non-SUSY BSM ideas, from new gauge bosons, quark compositeness, high mass resonances, extra dimensions and back holes, both in a model-independent and in a model-directed way, making sure no stone is left un-turned. Examples of a selection of results of such searches can be seen in figure~\ref{fig17} and figure~\ref{fig18} for the ATLAS and CMS collaborations, respectively. 

\begin{figure}[ht]
\begin{center}
\includegraphics[width=15cm]{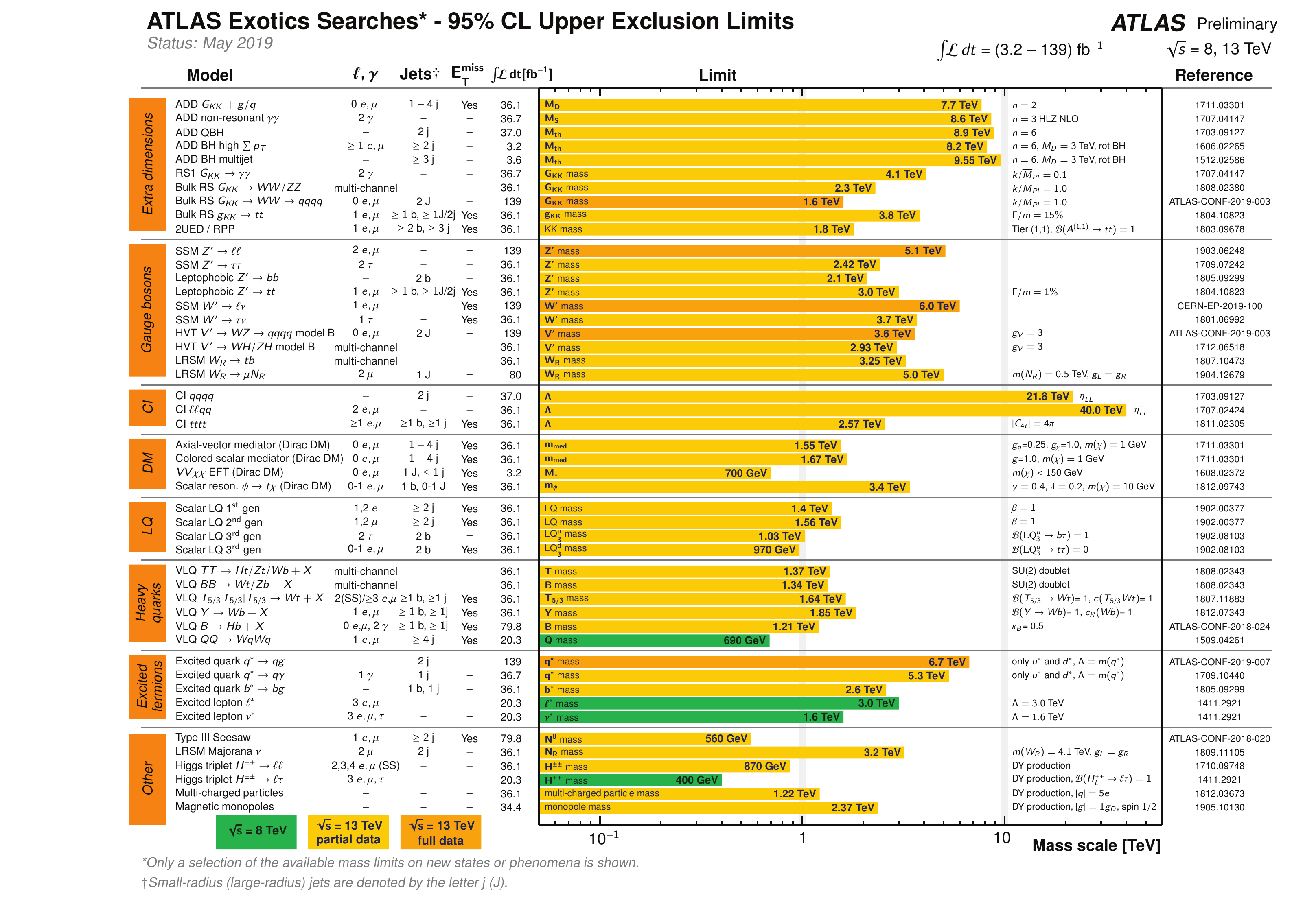}
\caption{A representative selection of available results from the ATLAS collaboration on searches for new phenomena other than SUSY. Green bands indicate 8 TeV data results; yellow (orange) bands indicate 13 TeV data results with partial (full) dataset.}
\label{fig17}
\end{center}
\end{figure}

\begin{figure}[ht]
\begin{center}
\includegraphics[width=15cm]{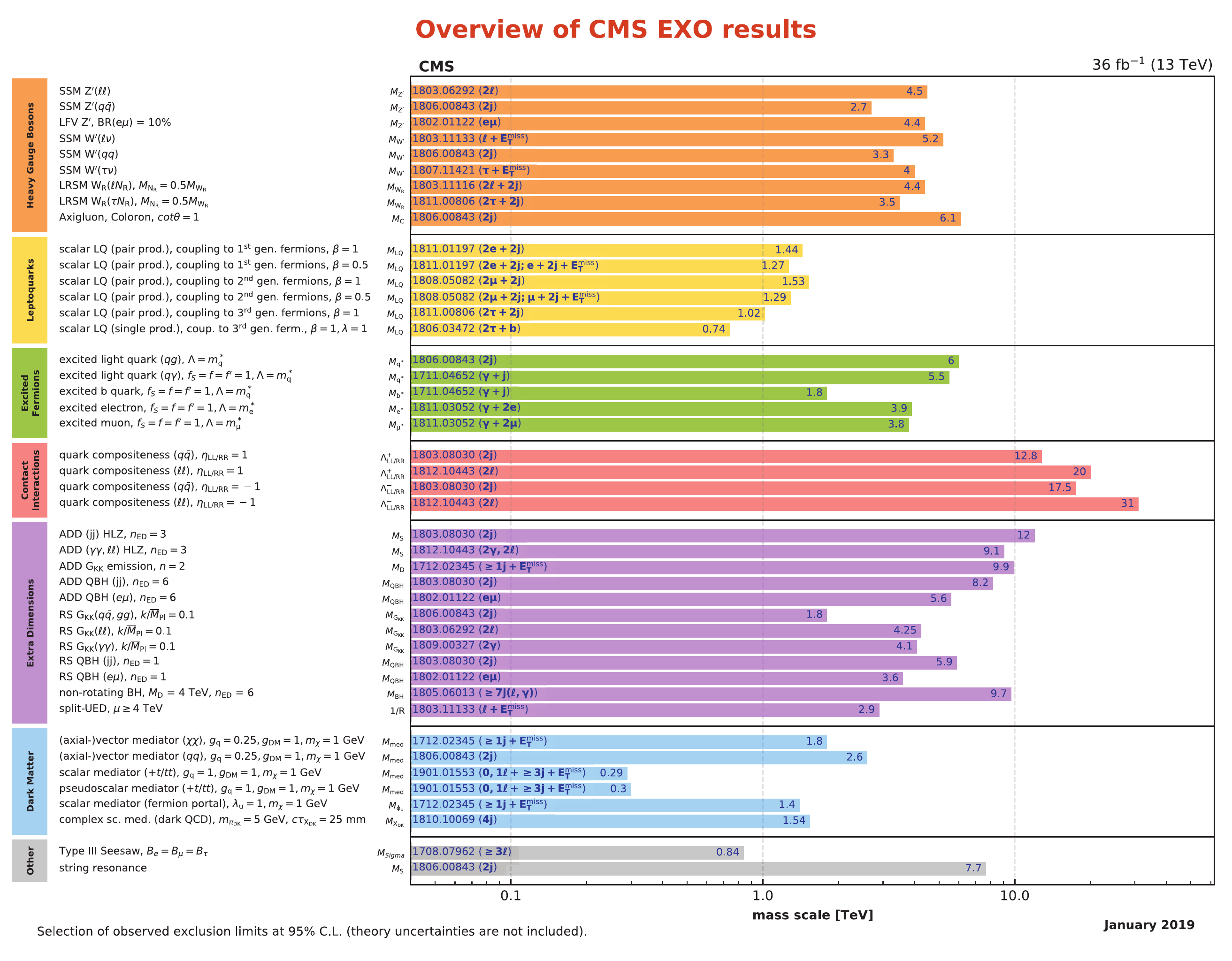}
\caption{Bar chart representing the mass scale reach of CMS BSM analyses using data collected in 2016 for a selected set of new physics phenomena. }
\label{fig18}
\end{center}
\end{figure}

In recent years, many novel techniques have been developed that rely on alternative methods to trigger and reconstruct events. One such example is the search for massive long-lived particles that would loose their kinetic energy and stop while traversing the detector. Such particles would give rise to energy deposits in the calorimeters or the muon systems, but have no associated hits in the tracking detectors (displaced particles). Alternatively, they would appear as tracks with no associated hits in the calorimeters or beyond (disappearing particles). These signatures pose a difficult experimental challenge and depend on modified object reconstruction techniques that do not assume the presence of a prompt vertex and rely on timing information for the energy deposits. Figure~\ref{fig19} shows a diagram of these hypothetical particles and their signatures and where they would decay depending on their lifetime.

\begin{figure}[ht]
\begin{center}
\includegraphics[width=7.5cm]{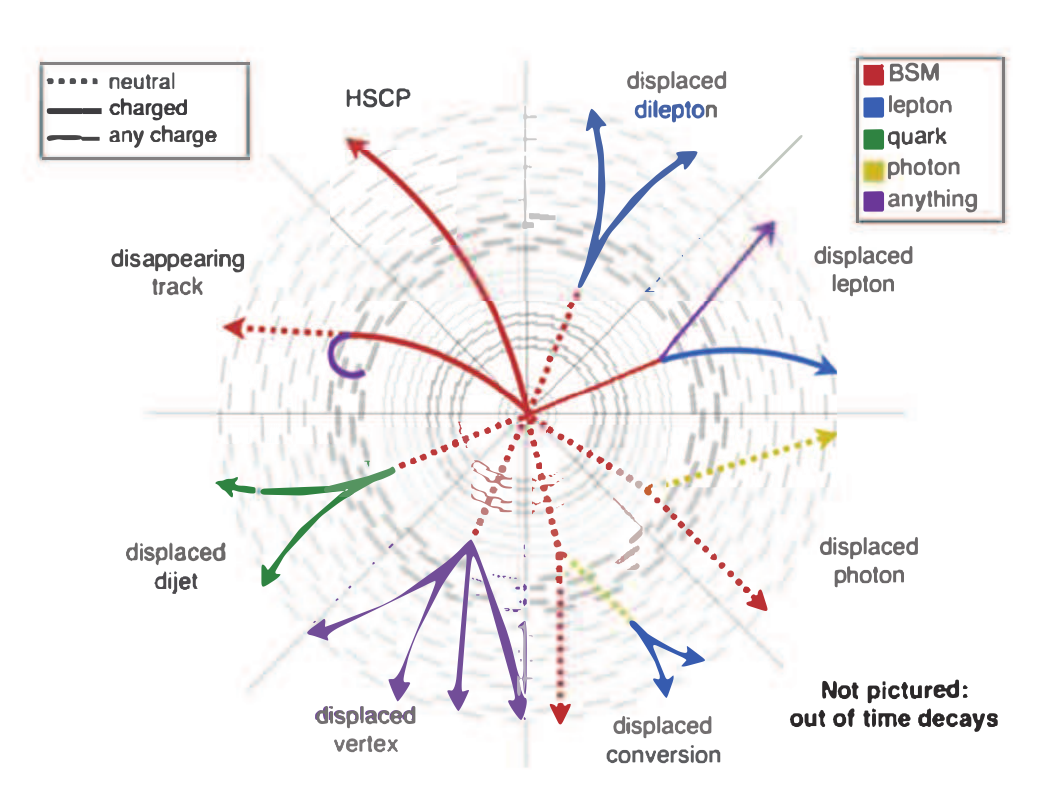}
\includegraphics[width=7.5cm]{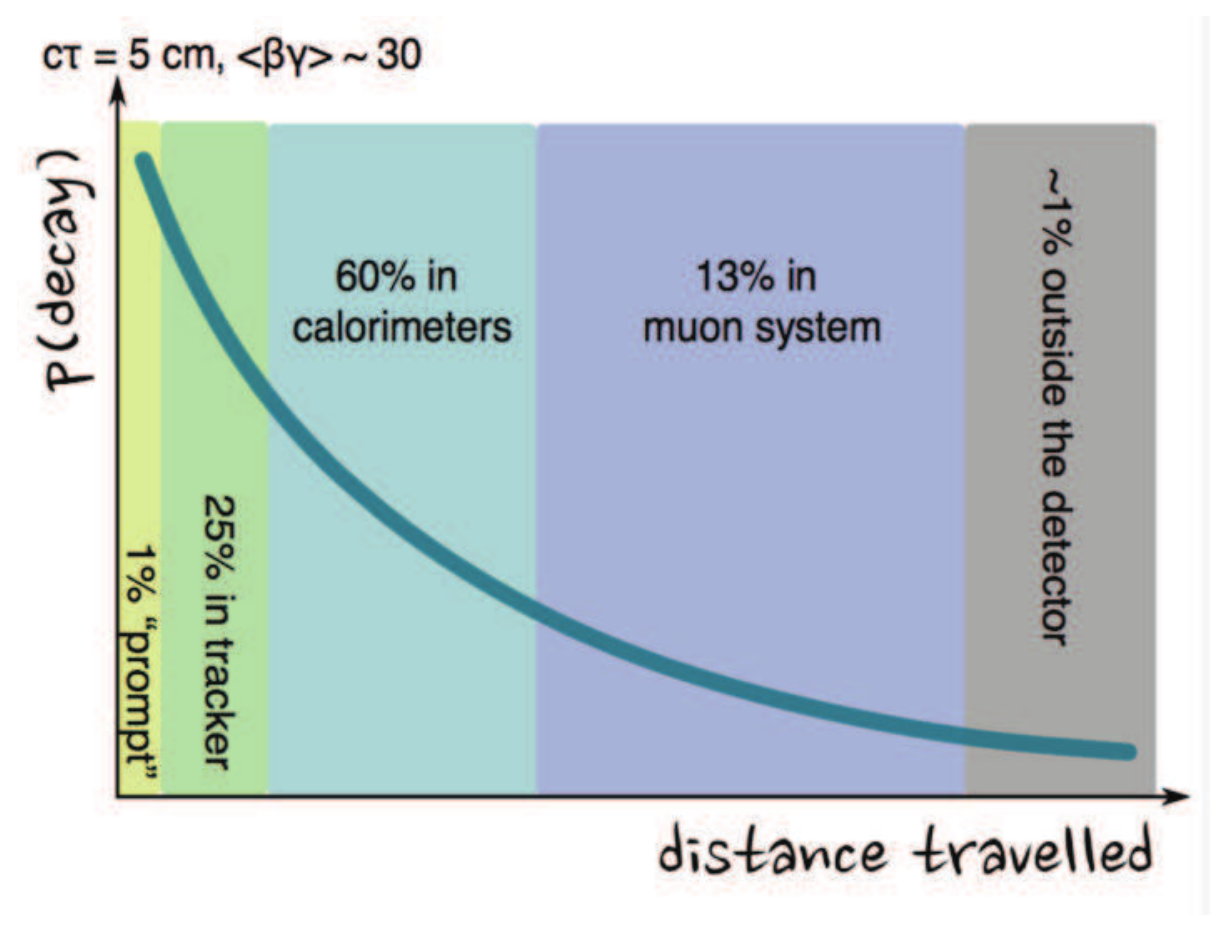}
\caption{Diagram of hypothetical long-lived particles and their signatures in the detectors (left). Distance travelled by a long-lived particle depending on its lifetime (right). }
\label{fig19}
\end{center}
\end{figure}

One example of such an analysis is the search for a single neutral long-lived particle $Z_d$, decaying hadronically, produced in association with a SM $Z$ boson which decays leptonically to electrons or muons~\cite{ATLAS-LL}. Such $Z_d$ particle would be produced in popular scenarios in hidden or dark sector models with additional $U(1)_d$ dark gauge symmetry~\cite{d1,d2} and would travel from a few centimeters to hundreds of meters, depending on the model. The ATLAS collaboration selected events with two opposite-sign isolated leptons and a jet that has no associated tracks and a timing inconsistent with out-of-time pileup and beam-induced backgrounds. The timing of the jet is obtained from the timing of its constituent calorimeter cells as measured by the tile calorimeter, relative to the expected time-of-flight from the bunch crossing to the cell~\cite{ATLAS-timing}. Figure~\ref{fig20} shows the reconstruction efficiency for the jet as a function of the transversal decay length and the obtained $95\%$ CL limits as a function of the decay length of $Z_d$. As can be seen, the efficiency is high beyond the volume of the tracker and maximum within the hadronic calorimeter. No significant excess of events is observed above the expected background, which allows the ATLAS collaboration to set limits on the production cross section of the $Z_d$ particle as a function of its mass for decay lengths from a few centimeters to one hundred meters.

\begin{figure}[ht]
\begin{center}
\includegraphics[width=7.5cm]{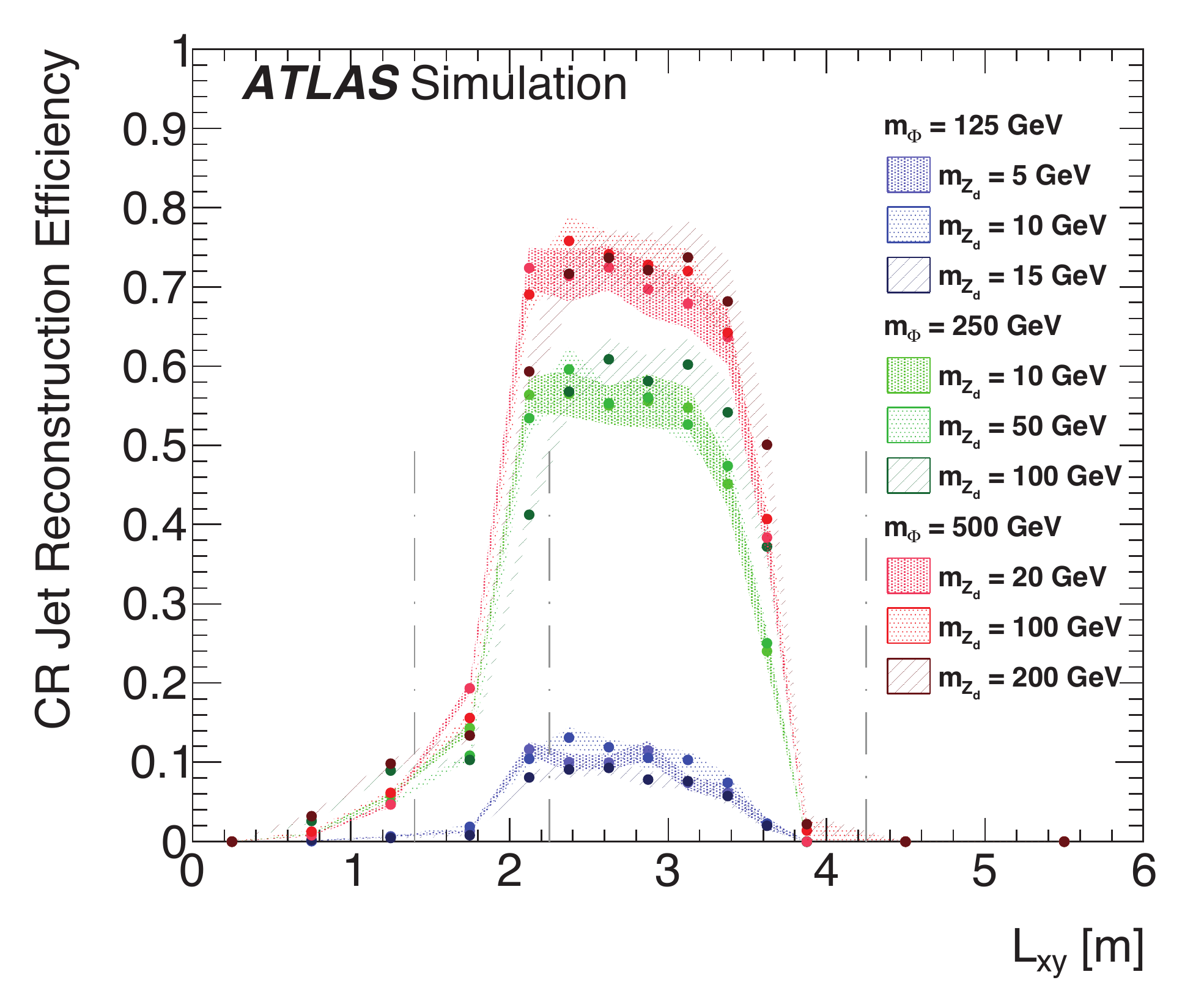}
\includegraphics[width=7.5cm]{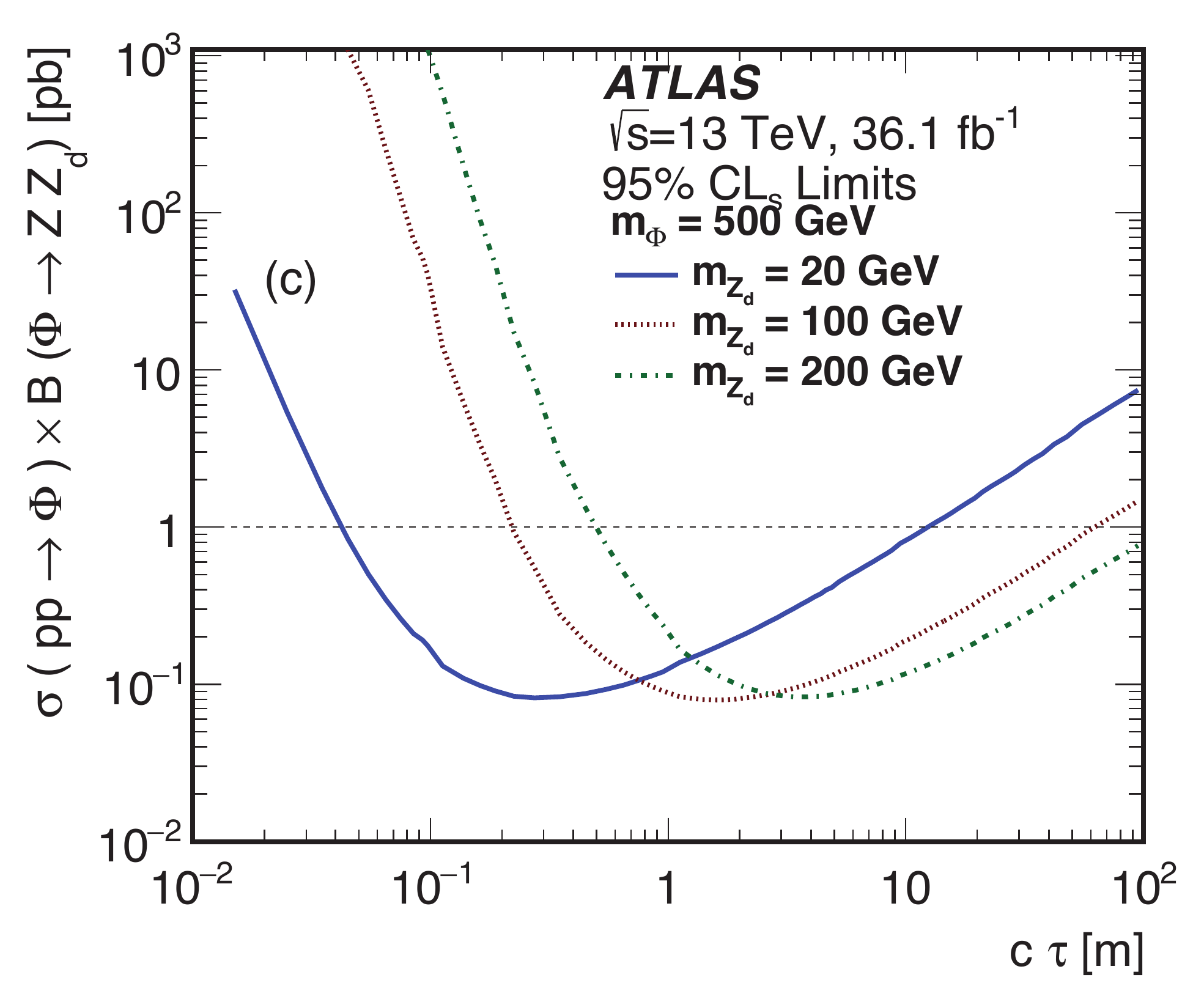}
\caption{Reconstruction efficiency for the jet as a function of the transversal decay length (left). $95\%$ CL limits as a function of the decay length of $Z_d$ (right).}
\label{fig20}
\end{center}
\end{figure}

The CMS collaboration also searched for the decay of heavy long-lived particles that come to rest in the detector, and whose decays would be visible during periods of time well separated from the pp collisions~\cite{CMS-LL}. For particles with lifetimes longer than tens of nanoseconds, their decays would be reconstructed as part of a separate event from their production. The search thus focuses on times when there are no proton bunches in the detector. Two cases are considered: a hadronic decay that would be detected as a large energy deposit in the calorimeters in the interval between collisions, and the case in which the particle decays into muons and appears as displaced muon tracks out of time with the collision. Dedicated triggers were deployed that are live only during specific time windows when the detector was quiet. Backgrounds arise from cosmic rays, beam halo, and detector noise, and are estimated from  control samples. Figure~\ref{fig20} (left) shows the difference in the time of the muon track between the upper and the lower hemisphere for data and estimated backgrounds. The data agrees well with events expected from cosmic rays. Figure~\ref{fig21} (right) shows an example of 95\% CL upper limits obtained from the calorimeter search in the neutralino vs. gluino mass plane, for lifetimes between 10 microseconds and 1000 seconds.

\begin{figure}[ht]
\begin{center}
\includegraphics[width=7.5cm]{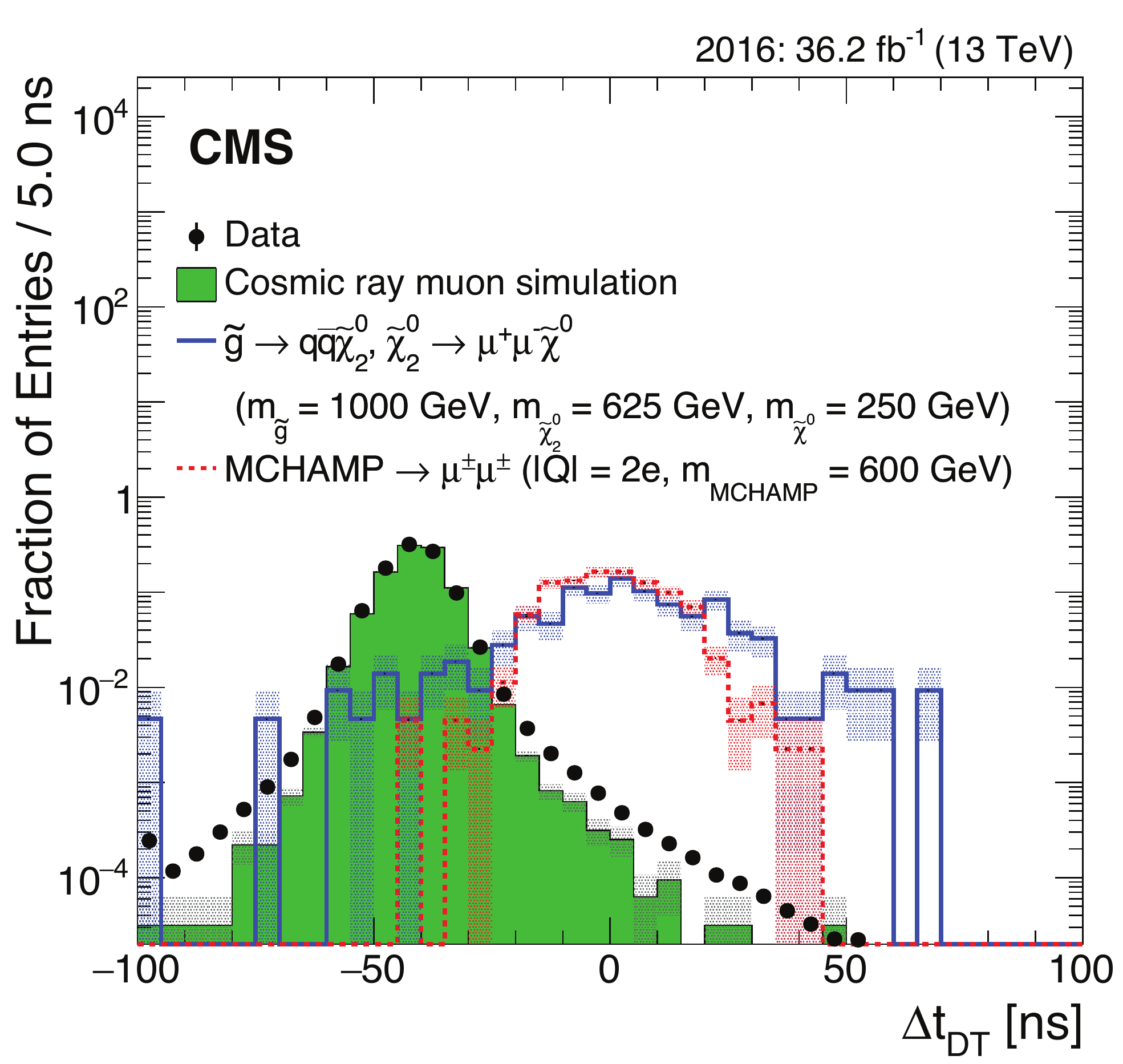}
\includegraphics[width=7.5cm]{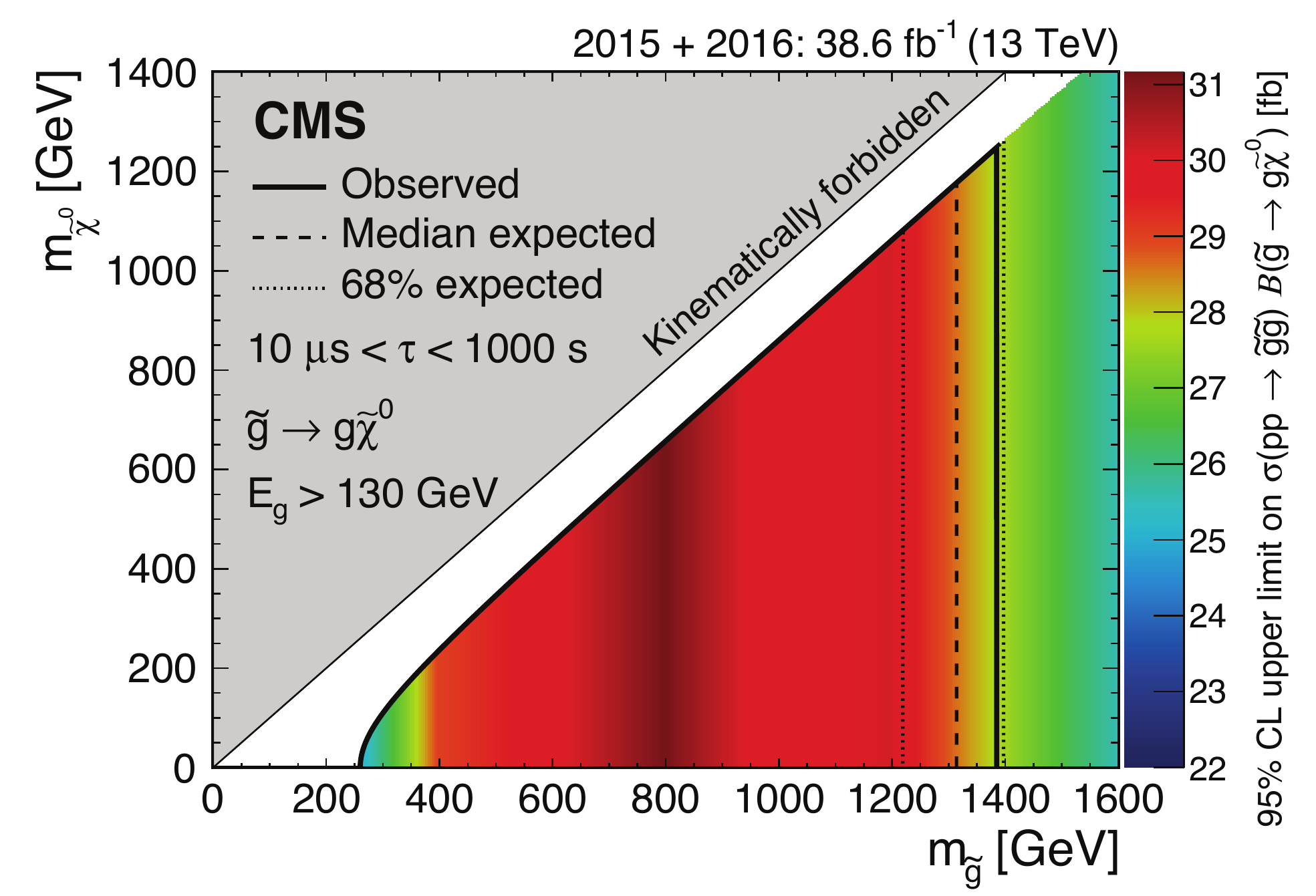}
\caption{Difference in the time of the muon track between the upper and the lower hemisphere for data and estimated backgrounds (left). Example of 95\% CL upper limits obtained from the calorimeter search (right).}
\label{fig21}
\end{center}
\end{figure}

\section{Dark Matter Searches}
Cosmological observations indicate that 85\% of the matter of the universe is Dark Matter (DM). However, there is no evidence yet for non-gravitational interactions between DM and SM particles. If such interaction existed, hadron colliders would offer a complementary strategy to look for non-gravitational DM interactions via the collision of SM particles at high energies. DM candidates are assumed to be weakly interacting and would leave no signal in the detectors. However, they could be identified by looking at the production of other particles decaying against them, giving rise to spectacular signatures in which jets or heavy particles would be seen recoiling against missing transverse energy. Early searches presented their results using effective field theory operators to describe the DM-SM interaction, allowing for direct comparison with non-collider searches in the contact interaction approximation. For cases in which the mediator of the DM-SM interaction is not very heavy, simplified models need to be used that include the particles and their BSM interactions and are valid at LHC energies~\cite{DM-theory}. These models are described by a small number of free parameters but make it harder to compare with direct and indirect detection experiments. Nevertheless, a rich phenomenology of DM searches at colliders has been pursued in the X+MET topology, where X includes single jets, photons, $W$ and $Z$ bosons, top and bottom quarks (single and pair). In all cases, control regions are used to understand the background contribution to the signal region and to ensure that spurious detector signals do not appear as fake missing transverse energy, mimicking the DM signal. Results are typically presented as exclusion plots in the DM vs. mediator mass plane, and the spin-independent (SI) or spin-dependent (SD) DM-nucleon or DM-proton cross section vs. DM mass.
Figure~\ref{fig22} shows examples in which LHC searches from ATLAS~\cite{ATLAS-DM} (left) and CMS~\cite{CMS-DM} (right) are compared to a selection of direct detection (DD) experiments. For the ATLAS result, the shaded areas are excluded both for the collider and the DD results. For the CMS result, the regions above the curves are excluded for DD experiments. The reinterpretation of the collider results in terms of a nucleon scattering cross section yields a higher sensitivity for lower masses than existing results from DD experiments, under the assumptions imposed by the model.

\begin{figure}[ht]
\begin{center}
\includegraphics[width=7.5cm]{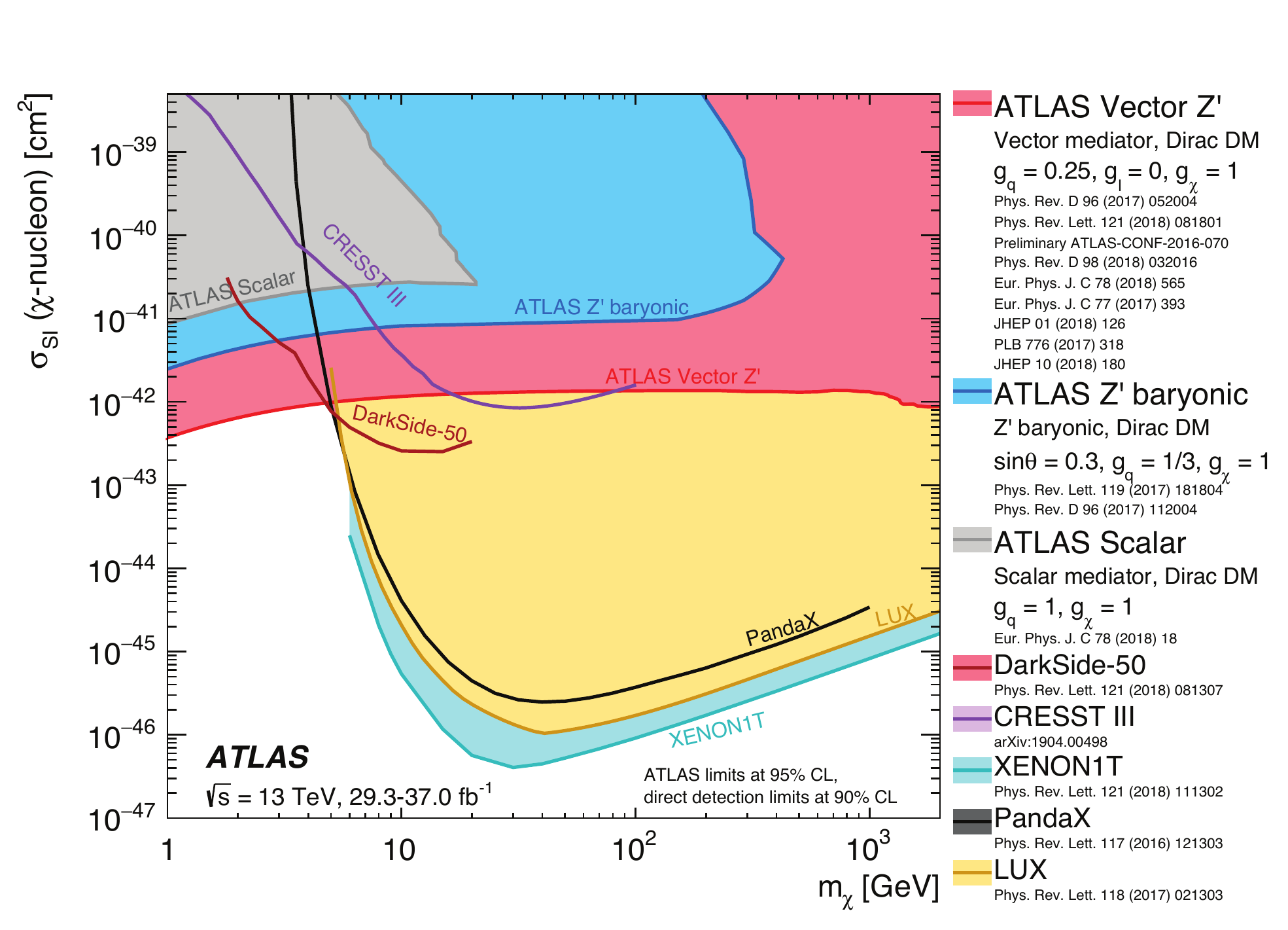}
\includegraphics[width=7.5cm]{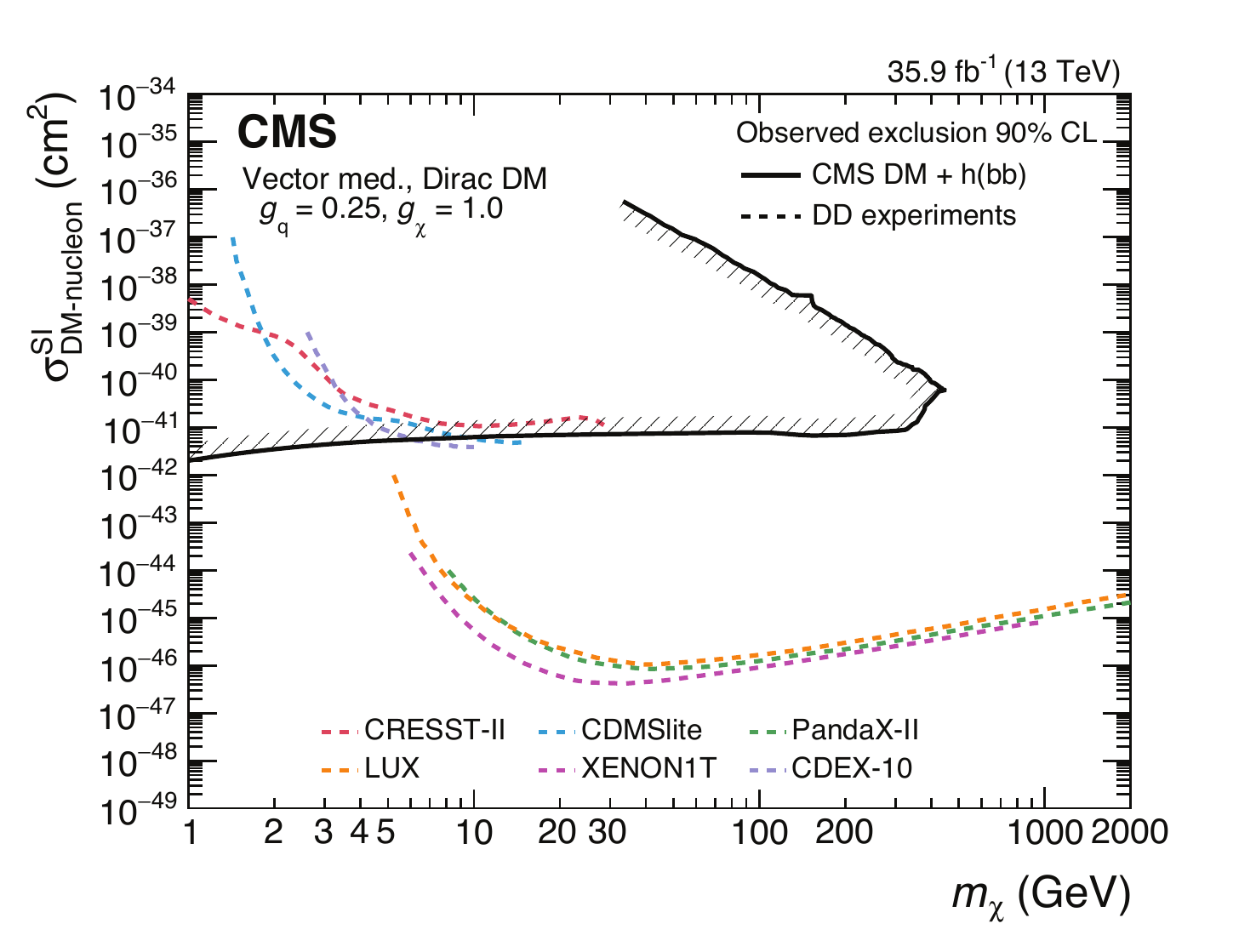}
\caption{Examples of Dark Matter searches in which the collider results are compared with results from direct detection experiments. Plots included as examples only and taken from~\cite{ATLAS-DM} (left) and CMS~\cite{CMS-DM} (right).}
\label{fig22}
\end{center}
\end{figure}

\clearpage
\section{The High-Luminosity LHC Era}
The LHC run plans for the next twenty years is well defined and summarized in figure~\ref{fig23}. We are currently in the Long Shutdown 2 (LS2) period after having collected about $150\;\rm fb^{-1}$ of data during Run 2. Run 3, currently scheduled for 2021-2023, will see an average of up to 80 simultaneous proton interactions per beam crossing, peak instantaneous luminosities of $2\times10^{34}\; {\rm cm^{-2}\;  s^{-1}} $ and an anticipated increase of the center of mass energy to $14\;\rm TeV$. The total integrated luminosity expected for Run 2 and 3 is $300\;\rm fb^{-1}$. After an extended long shutdown 3 (LS3), currently scheduled for 2024-2026, the proposed High Luminosity LHC (HL-LHC) Run 4 would begin in late 2026, with an average number of simultaneous pp collisions per beam crossing of 120, peak instantaneous luminosities of $5\times10^{34}\; {\rm cm^{-2}\;  s^{-1}} $, and a total integrated luminosity of $3000\;\rm fb^{-1}$. The ATLAS and CMS collaborations are planning a series of upgrades~\cite{UpgradeProposalATLAS, UpgradeProposalCMS, ATLAS-HLLHC, CMS-HLLHC} that will ensure the capabilities of the detector are matched to the running conditions expected from the LHC machine, while taking the opportunity to improve the performance and repair any problems uncovered during data-taking periods. The installation of the Phase 1 upgrades will be completed during LS2, and the Phase 2 upgrades are planned to coincide with LS3. 

\begin{figure}[ht]
\begin{center}
\includegraphics[width=16cm]{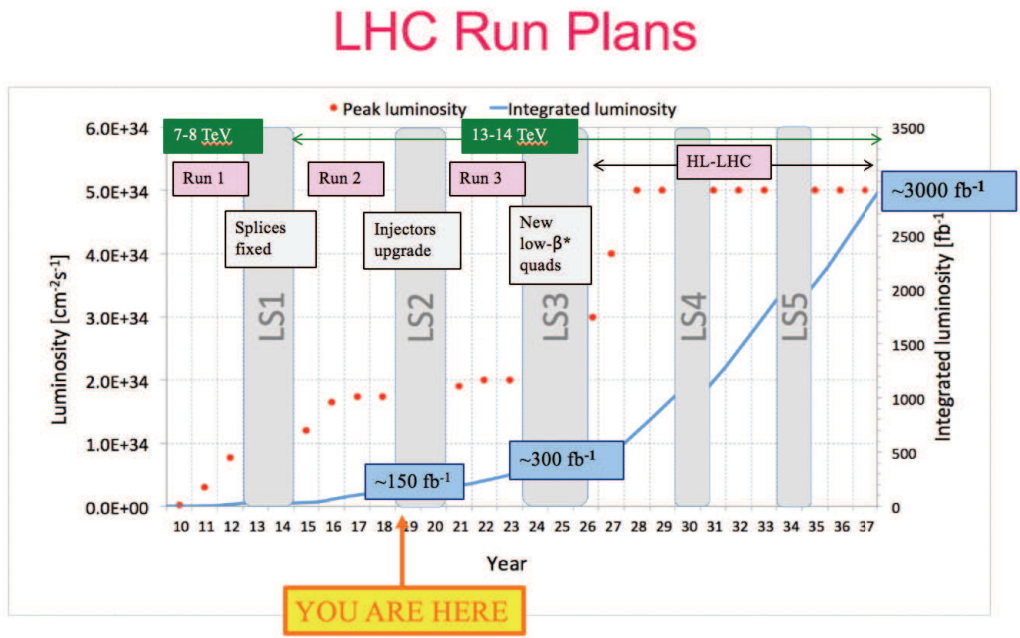}
\caption{The LHC run plans for the next twenty years, in which we expect to collect up to $3000\;\rm fb^{-1}$ of data, most of it at a center of mass energy of 14~TeV.}
\label{fig23}
\end{center}
\end{figure}
\clearpage

During the HL-LHC era, the large dataset will be delivered at the cost of having as many as 200 concurrent pp interactions every 25ns and large radiation doses. Both ATLAS and CMS are planning on significant changes to their detectors to maintain their performance in these challenging conditions. In particular, both detectors will add timing capabilities to cope with the increased pileup and preserve the lepton identification (via isolation), the b-tagging effectiveness (via primary vertex reconstruction and combinatorics) and energy measurement of the jets. Both experiments will also replace their entire tracking detectors, with ATLAS joining CMS in having an all-silicon tracker.  Both collaborations will increase the granularity and the coverage of their tracking volumes and reduce the material, which will allow them to preserve the reconstruction efficiency. Trigger capabilities require significant improvements to preserve the trigger thresholds and are implemented by installing higher bandwidth readout systems and adding fast tracking to the first trigger level. In addition, detectors that would be damaged by radiation are also being replaced by higher granularity, radiation-hard options. 

Physics Projection studies for the HL-LHC were prepared and submitted to the CERN Council Open Symposium on the Update of the European Strategy for Particle Physics~\cite{Granada}. From those studies, it is clear that the HL-LHC has an uncontested leadership in areas of direct searches for new particles, precision measurements of the Higgs boson, measurements of precision electroweak parameters and closure test ($W$ boson mass, top quark mass, Higgs boson mass), and some topics in rare B decays and other topics in B physics. 

One of the main goals of the HL-LHC studies is to measure the Higgs couplings to a precision close to the percent level.  Another goal is the measurement of the Higgs boson trilinear self-coupling $\gamma_{HHH}$ via the study of di-Higgs production. Figure~\ref{fig24}, taken from~\cite{WG2}, show that the Higgs couplings are expected to be measured with a precision that would be sensitive to new physics, and that the measurements of the Higgs trilinear interaction would provide constraints on the shape of the Higgs potential close to the minimum, verifying the electroweak symmetry breaking mechanism of the standard model.

\begin{figure}[ht]
\begin{center}
\includegraphics[width=16cm]{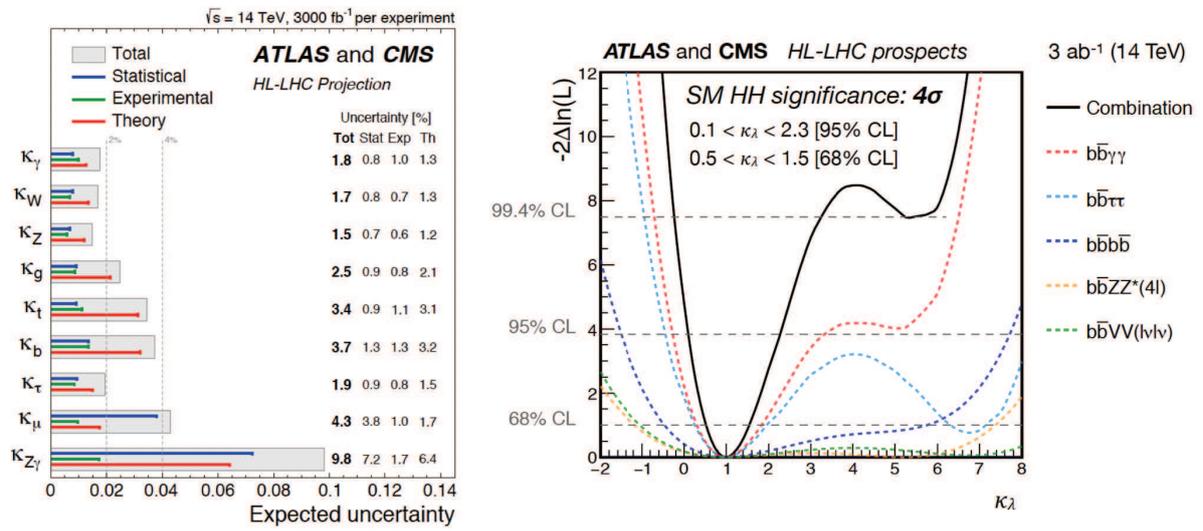}
\caption{Expected uncertainties on the coupling modifier parameters that are introduced to investigate potential deviations from the standard model prediction of the Higgs boson couplings to bosons and fermions (left). Minimum negative log-likelihood as a function of $\kappa_{\lambda}=\gamma_{HHH}/\gamma_{HHH}^{SM}$ (right).}
\label{fig24}
\end{center}
\end{figure}

The LHC results have confirmed the predictions of the standard model to unprecedented precision, and it is expected that the data collected during the HL-LHC era will extend the sensitivity to possible anomalies that might indicate the presence of new physics. Of particular interest are the studies of precision electroweak measurement and precision top quark physics, which can be combined together in a global fit of electroweak precision observables now that the Higgs Boson mass has been measured. This last input to the global fit of electroweak precision observables (EWPO) can be used to constrain new physics, a key goal of the HL-LHC physics program. Figure~\ref{fig25} shows the projections on the uncertainty of the top quark and the $W$ boson mass, as well as comparisons between the indirect constraints and current and projected measurements. What is apparent from the comparisons is that if the central values of the measured inputs were to remain unchanged, the expected improvement on their uncertainties would significantly increase the tension between the indirect determinations from the electroweak fit and the corresponding measurements. 

\begin{figure}[ht]
\begin{center}
\includegraphics[width=16cm]{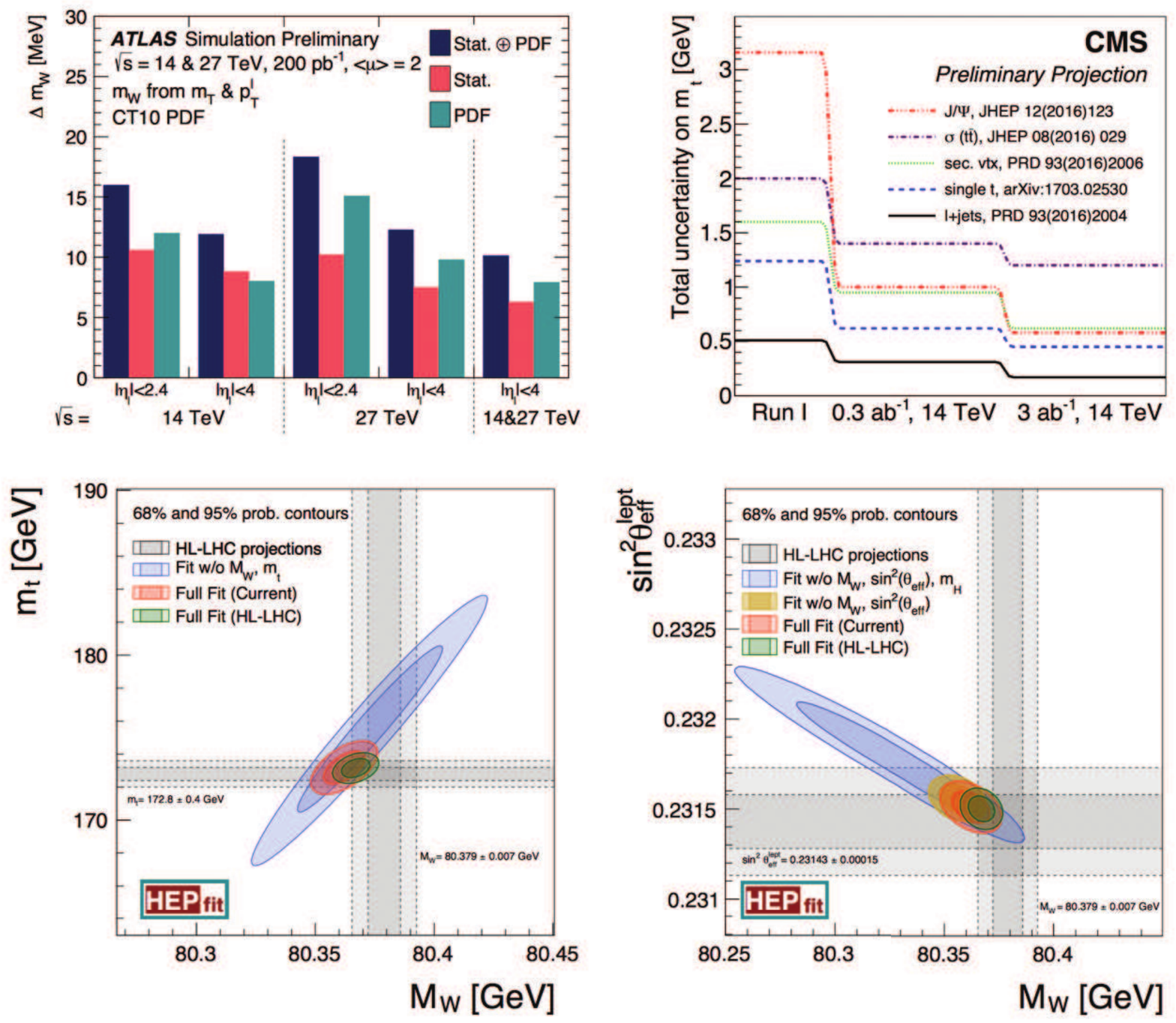}
\caption{Projections on the uncertainties for the measurements of the $W$ boson mass (top left), top quark mass (top right) and indirect constrains from fits to the electroweak precision observables compared with their measured values in the $W$ boson mass vs. top quark mass plane (bottom left) and vs. the effective weak mixing angle (bottom right).}
\label{fig25}
\end{center}
\end{figure}

The enormous amount of data expected during the HL-LHC era will open the door to precision multi-dimensional differential \ttbar cross section measurements that can be used as input to fits for parton distribution functions. The extended forward coverage that will be available with the upgraded ATLAS and CMS detectors will allow for fine-binned measurements in regions of phase space that were not previously accessible. These combined effects will result in unprecedented reductions on the uncertainties of the gluon parton distribution functions once the \ttbar data is incorporated in the fit, as can be seen in figure~\ref{fig26}. 

\begin{figure}[ht]
\begin{center}
\includegraphics[width=16cm]{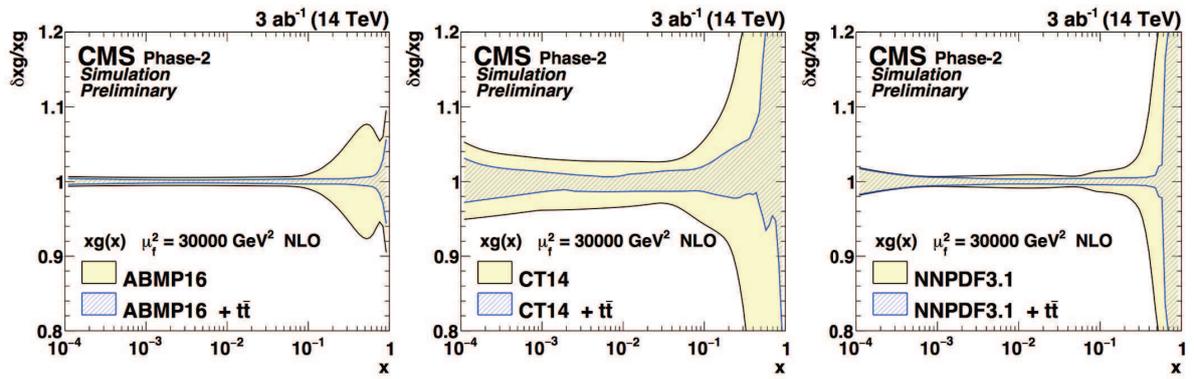}
\caption{Projections on the reduction of the relative gluon parton distribution function uncertainties after incorporating double-differential cross section \ttbar data to the fit.}
\label{fig26}
\end{center}
\end{figure}

\clearpage

\section{Outlook}
The LHC recently entered a two-year shutdown. We expect to double the data sample during Run 3, scheduled for 2021, with 10 times more data following in the HL-LHC era, currently scheduled for 2025-2035. The Higgs boson will continue to play a central role in the LHC physics program. The formidable precision era enabled by the HL-LHC will allow us to continue to probe the standard model predictions and look for cracks that might indicate the presence of new physics processes, even if their masses are above the LHC reach. Direct searches for beyond the standard model phenomena will continue to cover previously unexplored ground, with the collaborations pursuing both model-guided and model-independent searches to make sure all options are covered. Furthermore, new detector capabilities will allow us to search in previously unexplored regions like long-lived particles and very forward processes.

Exciting technical challenges lie ahead. If history is our guide, prior projections will be regularly surpassed by real results once the data is in hand and new techniques are developed, and surprises might be just around the corner. We have only collected $5\%$ of the data we expect from the LHC, and analyzed $1\%$ in most cases. We may not have seen an obvious sign of new physics in the data yet, however, what that implies is that we have to get cleverer and make sure we look in every corner and leave no stone unturned. Fun and exciting times lie ahead of us and there is no better time to join the quest.

\section*{Acknowledgements}

I wish to thank the ATLAS and CMS colleagues for their help in the preparation of these lectures. In particular, I would like to acknowledge the following individuals who gave me access to material that was ultimately incorporated into my presentations, some of which also made it to this summary: Jamie Antonelli, Doug Berry, Richard Cavanaugh, Matteo Cremonesi, Albert De Roeck, Sarah Eno, Frank Hartmann, Jim Hirschauer, Markus Klute, Andrey Korytov, Jeremy Mans, Corrinne Mills, Chris Palmer, Justin Pilot, Caterina Vernieri.

\end{document}